\documentclass[pra,twocolumn,showpacs,superscriptaddress,amssymb]{revtex4}

\usepackage{psfrag,graphicx}
\usepackage{dcolumn}
\usepackage{bm}
\usepackage{amsfonts,amssymb,amsmath}        

\usepackage{xcolor}

\newcommand{\be}{\begin{equation}}
\newcommand{\ee}{\end{equation}}
\newcommand{\bq}{\begin{eqnarray}}
\newcommand{\eq}{\end{eqnarray}}

\newcommand{\no}{\nonumber\\}

\newcommand{\rf}[1]{(\ref{#1})}
\newcommand{\SIGMA}{\mbox{\boldmath${\sigma}$}}
\newcommand{\PSI}{\mbox{\boldmath${\psi}$}}

\newcommand{\1}{1\!\!1}
\newcommand{\SSIGMA}{\mbox{\boldmath${\Sigma}$}}
\newcommand{\p}{{\bf p}}
\newcommand{\ket}[1]{\left |#1 \right\rangle}
\newcommand{\bra}[1]{\left \langle #1 \right |}
\begin{document}

\title{Seeing Majorana fermions in time-of-flight images of staggered spinless fermions coupled by $s$-wave pairing} 

\author{Jiannis K. Pachos}
\affiliation{School of Physics and Astronomy, University of Leeds, Leeds, LS2 9JT, United Kingdom}
\author{Emilio Alba}
\affiliation{Instituto de F\'{\i}sica Fundamental, IFF-CSIC, Calle Serrano 113b, Madrid 28006, Spain} 
\author{Ville Lahtinen}
\affiliation{Nordita, Royal Institute of Technology and Stockholm University, Roslagstullsbacken 23, SE-106 91 Stockholm, Sweden}
\affiliation{University of Amsterdam, Institute of Physics, Science Park 904, 1090 GL, Amsterdam, The Netherlands}
\author{Juan J. Garcia-Ripoll}
\affiliation{Instituto de F\'{\i}sica Fundamental, IFF-CSIC, Calle Serrano 113b, Madrid 28006, Spain}

\date{\today}


\begin{abstract}
The Chern number, $\nu$, as a topological invariant that identifies the winding of the ground state in the particle-hole space, is a definitive theoretical signature that determines whether a given superconducting system can support Majorana zero modes. Here we show that such a winding can be faithfully identified for any superconducting system ($p$-wave or $s$-wave with spin-orbit coupling) through a set of time-of-flight measurements, making it a diagnostic tool also in actual cold atom experiments. As an application, we specialize the measurement scheme for a chiral topological model of spinless fermions. The proposed model only requires the experimentally accessible $s$-wave pairing and staggered tunnelling that mimics spin-orbit coupling. By adiabatically connecting this model to Kitaev's honeycomb lattice model, we show that it gives rise to $\nu = \pm 1$ phases, where vortices bind Majorana fermions, and $\nu=\pm 2$ phases that emerge as the unique collective state of such vortices. Hence, the preparation of these phases and the detection of their Chern numbers provide an unambiguous signature for the presence of Majorana modes. Finally, we demonstrate that our detection procedure is resilient against most inaccuracies in experimental control parameters as well as finite temperature.
\end{abstract}

\maketitle

\section{Introduction}

Since the first theoretical proposal for realising Majorana modes -- zero energy quasiparticles that are their own anti-particles -- in solid state systems \cite{Kitaev01}, there has been a sustained research into a variety systems that might support them. This effort is partially motivated by the prospect of topological quantum computation \cite{Pachos12}. The general conditions for a fermionic system to support localised Majorana zero modes are understood: the spectrum should possess particle-hole symmetry and the ground state should exhibit suitable topologically non-trivial behaviour. Particle-hole symmetry implies that for a stationary state $\Psi^\dagger_E$ with energy $E$, there exists another state $\Psi_{-E}$ with energy $-E$. The suitable topological character of the ground state necessitates the presence of chiral edge states \cite{Hatsugai93}, which in turn imply that zero energy modes can be localised at the core of vortices \cite{Chamon1}. Due to particle-hole symmetry these $E=0$ modes satisfy the Majorana criterion $\Psi_0^\dagger=\Psi_0$. 

Particle-hole symmetry is an intrinsic property of superconducting fermionic systems. They can also exhibit the topological non-triviality required for Majorana modes when the pairing is either of $p$-wave type \cite{Read00} or the fermions in a more conventional $s$-wave superconductor are strongly spin-orbit coupled \cite{Fu08}. While recent experiments in solid state systems of latter type have yielded evidence supporting the existence of Majorana modes \cite{Mourik12,Das12,Churchill13}, loopholes remain \cite{Kells12,Stanescu12} and thus it is desirable to find other systems where Majorana modes could be unambiguously prepared and detected. An attractive platform are cold atoms trapped in optical lattices, where various directions have been taken: $p$-wave pairing could be induced either directly \cite{Gurarie07,Massignan10} (although hard experimentally \cite{Regal03,Guenter05}) or dissipatively \cite{Bardyn12}, the required spin-orbit interaction could be synthesised using several atomic states \cite{Sato09,Zhang08,AllStars}, or analogue one-dimensional superconducting wires could be directly realised \cite{Kraus12}. Here we take another approach, namely that of staggered spinless fermions. These can be realised with a single atomic species only, with the staggering giving rise to an effective pseudospin-orbit coupling. Thus when $s$-wave pairing is induced, one expects to find Majorana mode supporting phases. We will prove this by explicitly mapping our model to Kitaev's celebrated honeycomb model \cite{Kitaev06}, which in turn is adiabatically equivalent to the $p$-wave superconductor \cite{Yu08}. 

The ultimate goal is the experimental detection of Majorana modes. Like in the recent solid state experiments \cite{Mourik12,Das12,Churchill13}, also in optical lattices this has been proposed to be carried out by probing local densities \cite{Sato09, AllStars, Kraus12}. However, as the characteristic signals may arise also in non-topological phases \cite{Kells12}, it would be desirable to independently verify that the system is indeed in the correct topological phase. Theoretically non-interacting topological phases can be characterised by a topological number, such as the Chern number $\nu \in \mathbb{Z}$. Detecting this topological invariant would fully characterise the state of the system, with odd $\nu$ superconducting states supporting localised Majorana modes. Unfortunately, except for cases such as the off-diagonal conductivity in the quantum Hall effect \cite{Thouless}, it is in general not directly related to measurable quantities.

Here we provide such a connection by showing how to reproduce the Chern number of a general superfluid of fermionic atoms from time-of-flight images \cite{Spielman,Alba11}. Applying it to our model, we can robustly detect phases, both in the presence of finite temperature and of experimental imperfections, with Chern numbers $\nu=0$, $\pm1$ and $\pm2$. Due to the adiabatic connection to Kitaev's honeycomb model, we can immediately understand the nature of these phases. The $\nu=\pm1$ phases correspond to a regime where isolated vortices can bind interacting Majorana modes \cite{Lahtinen11}. The $\nu=\pm2$ phases, on the other hand, have been shown to emerge as a {\it unique} collective state of such Majorana modes bound to an underlying vortex lattice \cite{Lahtinen12}. While our detection scheme is applicable also to other experimental proposals, the detection of the $\nu=\pm2$ phases of our model would thus constitute an unambiguous global signature that Majorana modes {\it do exist} -- these phases emerge if and only if the model supports localised Majorana modes. Finally, we show that both the simulation of the superconducting model and the required time-of-flight measurements can be robustly implemented in state-of-the-art ultracold atom experiments \cite{Altman,Greiner,Foelling}.

This paper is organized as follows. In Section II we show how the Chern number for a superconducting system can be reproduced as a winding number of a vector whose components are obtained from physical observables. This construction is then generalized to staggered systems where we show the Chern number to be reproduced as the sum of physically observable winding numbers for each sublattice. In Section III we introduce a model of staggered spinless fermions and show that its rich phase diagram can be faithfully reproduced from the physically observable winding numbers. Analytic solution to the staggered model and its adiabatic connection to Kitaev's honeycomb lattice model are given in Appendices \ref{App_model} and \ref{App_honeycomb}, respectively. Finally, in Section IV we discuss the general implementation of the staggered model in optical lattices and outline a protocol for the experimental detection of the winding numbers. A quantitative analysis of the optical lattice implementation is left to Appendix \ref{App_implementation}.

\section{Chern number as an observable in topological superconductors} 

In this section we first explain how the Chern number of a translationally invariant topological superconductor can be computed as a physically observable winding number. Then we show that the winding number can be generalized to multi-component systems that arise in the presence of pseudospin degrees of  freedom, such as real spin, multiple orbitals or sublattices due to staggering, or several distinct species of atoms. We analytically demonstrate that the Chern number is reproduced as the sum of winding numbers for each pseudospin component. This decomposition is general and fails only when the pseudospin degrees of freedom are maximally entangled. 

In addition to the detection of the full Chern number, we will also show that its parity can be obtained from experimentally accessible density measurements. While not providing full characterization, this provides a simple method to distinguish between phases which can and can not support Majorana modes.

\subsection{The Chern as a winding number in a spinless system}

Formally, the Chern number, $\nu$, can be defined as the winding number of the projector onto the ground state~\cite{Thouless}. When the Bogoliubov-de Gennes Hamiltonian is a $2 \times 2$ matrix, i.e. the system is fully translationally invariant, it can always be written as $H(\p) \propto \mathbf{S}(\p)\cdot \SIGMA$ for some vector field $\mathbf{S}(\p)$. Here $\SIGMA$ denotes a vector of Pauli matrices. The Chern number, $\nu$, is then equivalent to the winding number 
\be \label{Chern}
\tilde \nu[{\bf S}] = {1 \over 4 \pi} \int_{BZ} {{\bf s}(\p)} \cdot \left({\partial {\bf s}(\p) \over \partial{p_x}} \times {\partial{\bf s}(\p) \over \partial{p_y}}\right) \mathrm{d}^2p \in \mathbb{Z},
\ee
which counts how many times the normalised vector $\mathbf{s}=\mathbf{S}/|\mathbf{S}|$ winds around the Bloch sphere in the particle-hole space as one spans the whole Brillouin zone~\cite{Thouless}. We can evaluate this quantity if we know the components of the vector field $\mathbf{S}(\p)$. These components are observables that can be obtained as the ground state expectation values
\be
\mathbf{S(\p)}=\bra{\Psi}\SSIGMA_\p\ket{\Psi}, \qquad \SSIGMA_\p = \PSI^\dagger_\mathbf{p} \SIGMA \PSI_\mathbf{p},
\label{eqn:Sp}
\ee
with the physical observables $\SSIGMA_\p$ being given in the basis $\PSI^\dagger_\mathbf{p} = (a^\dagger_{\p},a_{-\p})$ of the BdG Hamiltonian $H(\p)$:
\bq
	\Sigma^{x}_{\p} & = & a^\dagger_{\p} a^\dagger_{-\p} + a_{-\p}a_{\p}, \nonumber \\
	\Sigma^{y}_{\p} & = & -ia^\dagger_{p} a^\dagger_{-\p} + ia_{-\p}a_{\p}, \label{eqn:singleobs} \\
	\Sigma^{z}_{\p} & = & a^\dagger_{\p} a_{\p} - a_{-\p}a^\dagger_{-\p}. \nonumber
\eq
This set of observables are a basis for the single pseudospin Hamiltonian and constitute a natural extension of the operators which construct the winding number in the case of topological insulators~\cite{Alba11}.

While $S^z$ is experimentally readily obtained from density measurements $\Sigma^{z}_{\p}$, the experimental measurement of the operators $\Sigma^{x}_{\p}$ and $\Sigma^{y}_{\p}$ is challenging, since they violate a superselection rule: the number of particles. However, one can in general go around this by mapping them to experimentally accessible operator $\Sigma^{z}_{\p}$ with suitable rotations on the state. This can typically be achieved by using operators present in the Hamiltonian (such as $\Sigma^{x,y}_{\p}$ themselves). We will later illustrate with a particular example how this could be performed in an optical lattice experiment.

\subsection{Winding numbers for the multi-component case}

To generalise the construction of the Chern number as a physically observable winding number to a system with $m$-site unit cell (or more generally, $m$ degrees of freedom giving $2^m$ dimensional Hilbert space per unit cell), we define an independent vector field $\mathbf{S}_{(i)}(\p)=\bra{\Psi}\SSIGMA_{(i),\p}\ket{\Psi}$ for each of the sublattices, $i=1,\ldots,m$. The corresponding sublattice observables $\SSIGMA_{(i),\p} = \PSI^\dagger_{(i),\mathbf{p}} \SIGMA \PSI_{(i),\mathbf{p}}$ are explicitly given by
\bq
	\Sigma^{x}_{(i),\p} & = & a^\dagger_{(i),\p} a^\dagger_{(i),-\p} + a_{(i),-\p}a_{(i),\p}, \nonumber \\
	\Sigma^{y}_{(i),\p} & = & -ia^\dagger_{(i),\p} a^\dagger_{(i),-\p} + ia_{(i),-\p}a_{(i),\p}, \label{eqn:manyobs} \\
	\Sigma^{z}_{(i),\p} & = & a^\dagger_{(i),\p} a_{(i),\p} - a_{(i),-\p}a^\dagger_{(i),-\p}. \nonumber
\eq
We now show how to construct, out of these observables, a quantity that (i) is an integer, (ii) is defined in terms of measurable quantities and (iii) reproduces the Chern number in the zero temperature limit. Substituting each set of sublattice observables into \rf{Chern}, we can construct $m$ winding numbers $\tilde{\nu}_{(i)}=\tilde \nu[\mathbf{S}_{(i)}]$, $i=1,...,m$, with the total winding number being defined as their sum
\be \label{winding}
	\tilde{\nu} = \sum_{i=1}^m \tilde{\nu}_{(i)}.
\ee
By construction, this quantity satisfies properties (i) and (ii) as listed above. To satisfy (iii) we present the following argument for reproducing the Chern number in terms of sublattice winding numbers. A more formal and general proof is presented in a follow up work \cite{Pachos13}.

\subsubsection{Proof for Chern number decomposition in terms of sublattice winding numbers}

The ground state of our model can in general be Schmidt decomposed as
\bq \label{GS_Schmidt}
  \ket{\Psi(\mathbf{p})} & = & \cos[\theta(\mathbf{p})] \ket{\phi^+_w(\mathbf{p})} \ket{\phi^+_b(\mathbf{p})} + \\
  \ & \ &\,\,\,\,\,\,\,\, \sin[\theta(\mathbf{p})] \ket{\phi^-_w(\mathbf{p})} \ket{\phi^-_b(\mathbf{p})},\nonumber
\eq
where $\cos[\theta(\mathbf{p})] \geq 0$ and $\sin[\theta(\mathbf{p})] \geq 0$, are the postive weights ($\theta \in [0,\pi/2]$) of the Schmidt decomposition and the orthonormal and momentum-dependent states $\{\ket{\phi^+_{b}},\ket{\phi^-_{b}}\}$ ($\{\ket{\phi^+_{w}},\ket{\phi^-_{w}}\}$) live only on the black (white) sublattice. When the states $\ket{\phi^+_{(i)}}$ are viewed as ground states of a two-dimensional Hamiltonian $H_{(i)}=(1+\ket{\phi^+_{(i)}}\bra{\phi^+_{(i)}})/2$, we associate a vector $\mathbf{S}_{(i)}$ to them through $H_{(i)} \propto \mathbf{S}_{(i)} \cdot \SIGMA$. It is then straighforward to verify that 
\be
  S_{(i)}^\alpha = \bra{\Psi} \Sigma^{\alpha}_{(i),\p} \ket{\Psi}   = T \bra{\phi^+_{(i)}} \Sigma^{\alpha}_{(i),\p} \ket{\phi^+_{(i)}}  = T  s_{(i)}^\alpha,
\ee
where we defined $T =  \cos^2\theta - \sin^2\theta$. The orthonormality of the states $\ket{\phi^\pm_{(i)}}$ gives
\begin{equation}
  |\textbf{S}_{(i)}| = |T| = |\cos^2\theta - \sin^2\theta|,
\end{equation}
which means that the norms of vectors $\textbf{S}_{(i)}$ are equal and provide a physically observable measure of the entanglement between the sublattices. For $\theta=0$ or $\pi/2 $ they are unentangled, while for $\theta=\pi/4 $ they are maximally entangled. In the latter case $|\textbf{S}_{(i)}|$ vanishes and the decomposition can no longer be described in terms of physically observable vectors $\textbf{S}_{(i)}$ associated with each sublattice. Assuming this is not the case, i.e. $\theta \neq \pi/4$ for all momenta, we can associate a winding number \rf{Chern} to each vector in the same way as in the spinless case.

The Chern number can be decomposed into a sum of these winding numbers as follows. It can be formally given as the Berry phase of the ground state along the edge of Brillouin zone
\begin{equation}
  \nu = \frac{1}{2\pi i}\oint_{\partial BZ} \langle \Psi| \nabla |\Psi\rangle \cdot \mathrm{d} \textbf{p}.
\end{equation}
Substituting the Schmidt decomposed ground state \rf{GS_Schmidt} into this expression and using the normalization of the state, we obtain
\begin{equation}
  \nu = \sum_{i} \frac{1}{2\pi i} \oint_{\partial BZ} T \bra{\phi^+_{(i)}} \nabla \ket{\phi^+_{(i)}} \cdot \mathrm{d}\textbf{p}.
\end{equation}
Without loss of generality we assume that $T>0$ for all momenta. Then one finds, up to a vanishing additive integral, that 
\be
  \oint T \bra{\phi} \nabla \ket{\phi} \cdot \mathrm{d}\textbf{p} = \oint \left( \bra{\phi}\sqrt{T}\right) \nabla \left(\sqrt{T}\ket{\phi}\right) \cdot \mathrm{d}\textbf{p}. \nonumber
\ee
As $\sqrt{T}$ plays only the role of a scaling of the normalized Bloch vector $\ket{\phi}$, the winding number on the right hand side remains invariant if we take $T \to 1$.  We can thus define sublattice ``Chern numbers'' as $\tilde{\nu}_{(i)} = \frac{1}{2\pi i} \oint_{\partial BZ} \bra{\phi^+_{(i)}} \nabla \ket{\phi^+_{(i)}} \cdot \mathrm{d}\textbf{p}$ in terms of which the Chern number of the ground state is additive. Realizing that each $\tilde{\nu}_{(i)}$ can be evaluated as the the winding number \rf{Chern} of the corresponding normalized vectors $\textbf{s}_{(i)}$, we arrive at the conclusion \rf{winding} that the Chern number of the full ground state can be obtained as the sum of winding numbers associated with physical observables on each sublattice.

For this decomposition to make sense, we assumed that the vectors $\textbf{S}_{(i)}$  can be robustly determined, i.e. that they have a finite norm. This requirement thus provides a physical constraint for the detection of the Chern number: The Chern number is reproduced as the sum of the sublattice winding numbers only when the sublattices are not maximally entagled. As the entanglement given by the norm $|\textbf{S}_{(i)}|$ is also a physical observable, it can be used in the experiments as a measure of reliability of the characterization provided by the winding number \rf{winding}. We will numerically verify in the next section that the decomposition indeed fails only in the maximal entanglement limit.




\subsection{Chern number parity from density measurements}

While the Chern number can be obtained by using the full set of observables \rf{eqn:manyobs}, for practical purposes a coarser classification of the phases can be sufficient. For instance, to distinguish between phases that support localized Majorana modes (odd $\nu$) from those that do not (even $\nu$), it is sufficient to know only the parity of the Chern number. Or to classify all the topological phases up to their chiralities, the knowledge of $|\nu|$ is sufficient. Remarkably, both can be obtained from $\Sigma^z_{(i)}$ measurements that are directly experimentally accessible. 

Let us consider first the properties of the spinless case. Due to the presence of both translational and particle-hole symmetries the surface ${\bf S}(\p)$ has always the topology of a torus and it is always symmetric around the $z$-axis. This means that by just counting the extremal and saddle points of the $S^z(\textbf{p})$, we can infer whether the surface ${\bf S}(\p)$ encloses the origin or not. The key observation is that $\tilde\nu\neq 0$ is possible only if it does. The parity of the winding number $|\nu| (\textrm{mod} \,2)$ can thus be obtained using the following simple protocol: (i) Find the null-gradient-points (local maxima and minima and saddle points) of the $S^z$ distribution in the Brillouin zone, and (ii) assign $|\tilde\nu|=0 (1)$ if the number of such points with $S^z>0$ is even (odd). Phases with $|\nu|=0 (1)$ will correspond to phases with even (odd) Chern numbers.

In a system with $m$ components one has $m$ winding numbers $\tilde{\nu}_{(i)}[{\bf S}_{(i)}]$ whose parities can be independently obtained using the same protocal as above. This allows for a richer characterization of the phases beyond just the Chern number parity. In fact, when we apply in the next section the parity measurements to a particular example, we find that the absolute value of the Chern number can be consistently obtained as the sum of the sublattice winding parities, i.e. that $|\nu| = \sum_i |\tilde{\nu}_{(i)}|$. We postulate that this is a general property, which allows for the full characterization of different types of topological phases in multi-component systems based on the experimentally accessible density measurements only.

\section{Case study: Staggered spinless fermions with $s$-wave pairing}

In this section we demonstrate our detection scheme for the Chern number in the context of a particular model. First we introduce a model of staggered spinless fermions whose phase diagram contains topological phase characterized by Chern numbers $\nu=0,\pm 1$ and $\pm 2$. We briefly discuss its adiabatic connection to Kitaev's honeycomb model (details given in Appendix \ref{App_honeycomb}) and the way this connection allows the model to exhibit collective signatures of Majorana modes. In the second part we demonstrate that the phase diagram of the model can be robustly captured using the detection methods described in Section II.

\subsection{The model}

\begin{figure}[t]
\includegraphics[width=\linewidth]{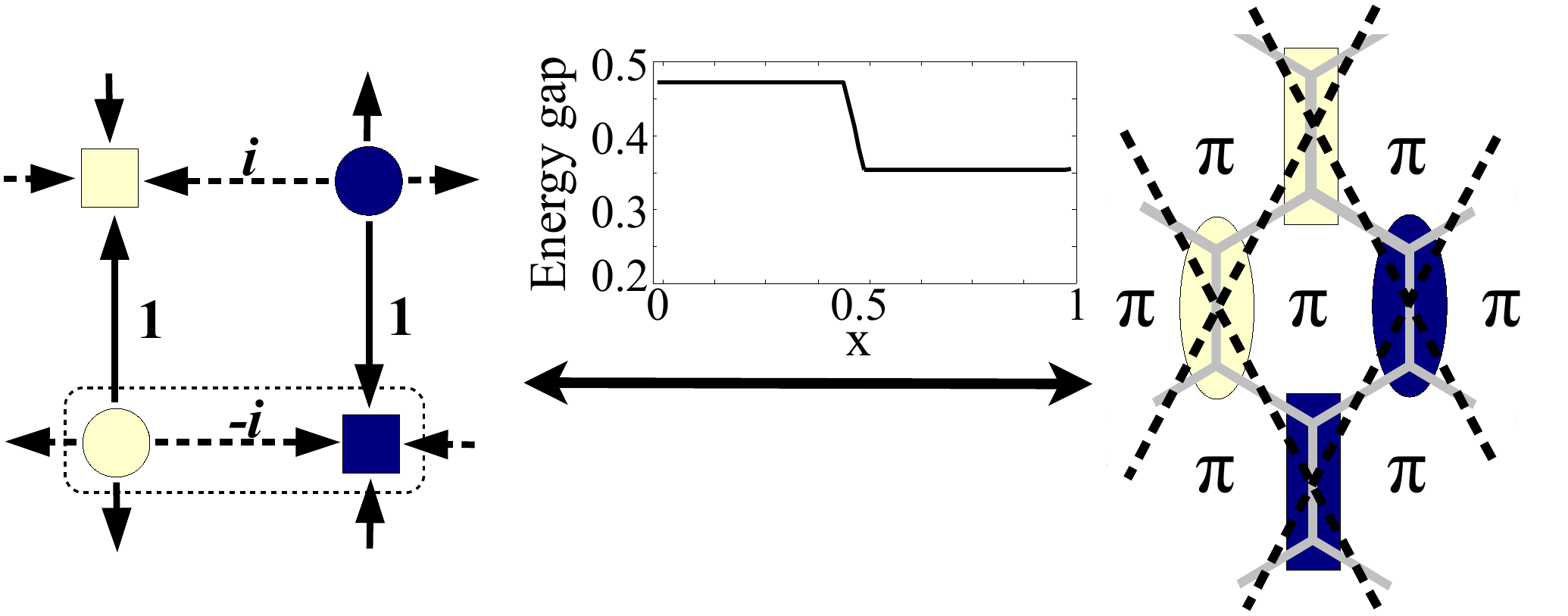}
\caption{
{\it Left:} Staggered topological superconductor with $s$-wave pairing on a square lattice~\rf{H2}. The numbers denote the relative phases of the tunnelling amplitudes, while the black (white) sites experience a chemical potential $\mu_b=\mu+\delta$ ($\mu_w=\mu-\delta$). Circles and squares denote the underlying distinct, but fixed internal atomic states that facilitate the optical lattice implementation. The dashed box denotes the two site ``magnetic'' unit cell.
{\it Right:} When Kitaev's honeycomb model with $\pi$-flux vortex per plaquette is written in the basis of complex fermions, the vertical links become the sites of a square lattice, with the fermions subject to a staggered chemical potential. As detailed in Appendix \ref{App_honeycomb}, a linear interpolation $x H + (x-1) H_\text{HC}$ for $x\in[0,1]$ shows that our model~\rf{H2} can be adiabatically connected to the honeycomb model with Hamiltonian $H_\text{HC}$.
}\label{lattice}
\end{figure}

Our model is defined for spinless fermions on a square lattice and combines staggered complex hopping with a uniform superconducting $s$-wave interaction. The Hamiltonian is
\bq \label{H2}
H & = & \sum_{\bf j} \Big[ \mu^{}_{\bf j} a^\dagger_{\bf j} a^{}_{\bf j} 
+ it(-1)^{j_x} a_{\bf j}^\dagger a^{}_{{\bf j}+{\hat{\bf x}}}+ta_{\bf 
j}^\dagger a^{}_{{\bf j}+\hat{\bf y}} \nonumber\\
& \ &\,\,\,\,\,\,\,\,\,\,\,\, + \Delta\big(a_{\bf j}^\dagger a_{{\bf 
j}+\hat {\bf x}}^\dagger + a_{\bf j}^\dagger a_{{\bf j}+\hat{\bf 
y}}^\dagger\big)\Big] +\mathrm{H.c.},
\eq
where $a_{\bf j}^\dagger$ creates a fermion at site ${\bf j}=(j_x,j_y)$, the tunnelling amplitude $t$ and the pairing potential $\Delta$ are both real and the chemical potential $\mu_{\bf j}=\mu+(-1)^{j_x}\delta$ is staggered by the detuning $\delta$. Translational symmetry is broken along the $x$-direction with the ``magnetic'' unit cell consisting of two adjacent sites with detuned chemical potentials, as shown in Fig.~\ref{lattice}. Inspired by the Kogut-Susskind fermions~\cite{Susskind,Maraner} we interpret this lattice degree of freedom as a ``pseudospin'' $\tau\in\{b,w\}$ of the fermions $a_{\tau,{\bf j}}^\dagger$. The Hamiltonian \rf{H2} can thus be viewed as an effective pseudospin-orbit coupled system: Tunnelling along the $x(y)$-direction changes (conserves) the pseudospin state, which effectively realises an anisotropic Rashba type spin-orbit coupling, while the chemical potential detuning plays the role of a Zeeman term. Thus, by adding $s$-wave pairing, one expects to find Majorana mode supporting topological phases \cite{Alicea10}.

\begin{figure}[t]
\includegraphics[width=\linewidth]{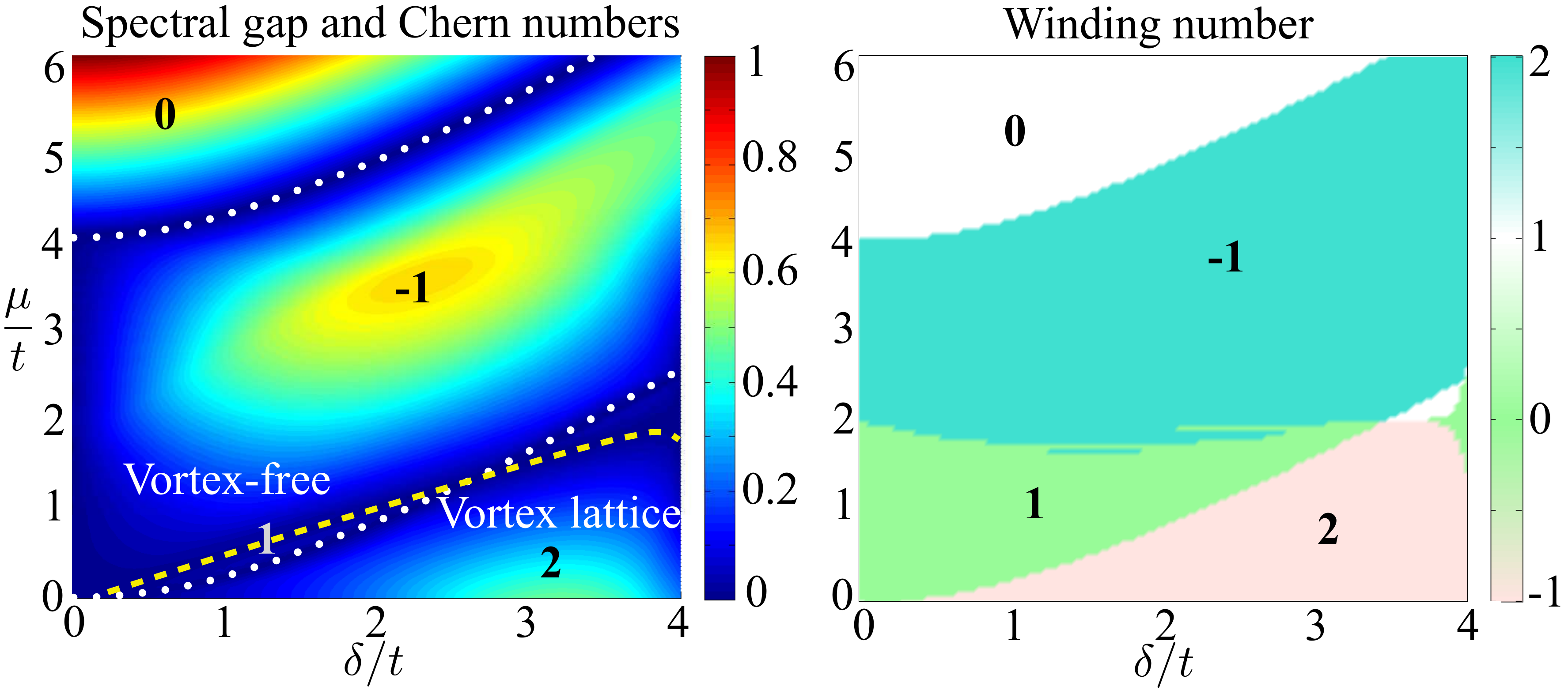}
\caption{\label{PD} {\em Left:} The phase diagram of \rf{H2} as a function of the overall chemical potential $\mu$ and its detuning $\delta$. Colour encodes the magnitude of the fermionic spectral gap, the dashed lines show the phase boundaries at which the gap closes. The Chern number $\nu$ for each phase is also shown. The phase diagram is symmetric with respect to $\mu  \to -\mu$, while for $\delta \to -\delta$ all the Chern numbers become time reversed ($\nu \to -\nu$). The regions $\mu \lesssim \delta/2$ ($\mu \gtrsim \delta/2$) can be identified with the honeycomb model in the presence (absence) of a vortex lattice (see Appendix \ref{App_honeycomb}). {\em Right:} The total winding number $\tilde{\nu}$, \rf{winding}, (encoded in colour), as obtained from the observables \rf{eqn:manyobs}. It shows perfect agreement with the Chern number except in regions where sublattices are close to being maximally entangled (see Fig.~\ref{PD_ent}). Both plots are for $\Delta/t=2$. }
\end{figure}

To verify this, we solve \rf{H2} by Fourier transforming it with respect to the magnetic unit cell. Writing it subsequently in the particle-hole basis $\PSI^\dagger_\p=(a^\dagger_{b,\p}, a^\dagger_{w,\p},a_{b,-\p}, a_{w,-\p})$, we obtain the quadratic Hamiltonian $H = \int_{BZ} \PSI^\dagger_\p H(\p) \PSI_\p \mathrm{d}^2p$,
where the Brillouin zone (BZ) spans $p_{x} \in [0,\pi]$ and $p_{y} \in [0,2\pi]$, and the Bloch Hamiltonian $H(\p)$ is a $4 \times 4$ matrix. From the analytic solution presented in Appendix \ref{App_model}, we obtain the phase diagram shown in Fig.~\ref{PD}. We find that by varying only the chemical potentials we can move between a variety of extended topological phases with Chern numbers $\nu=0, \pm 1$ and  $\pm 2$. 

\begin{figure}[t]
\includegraphics[width=5cm]{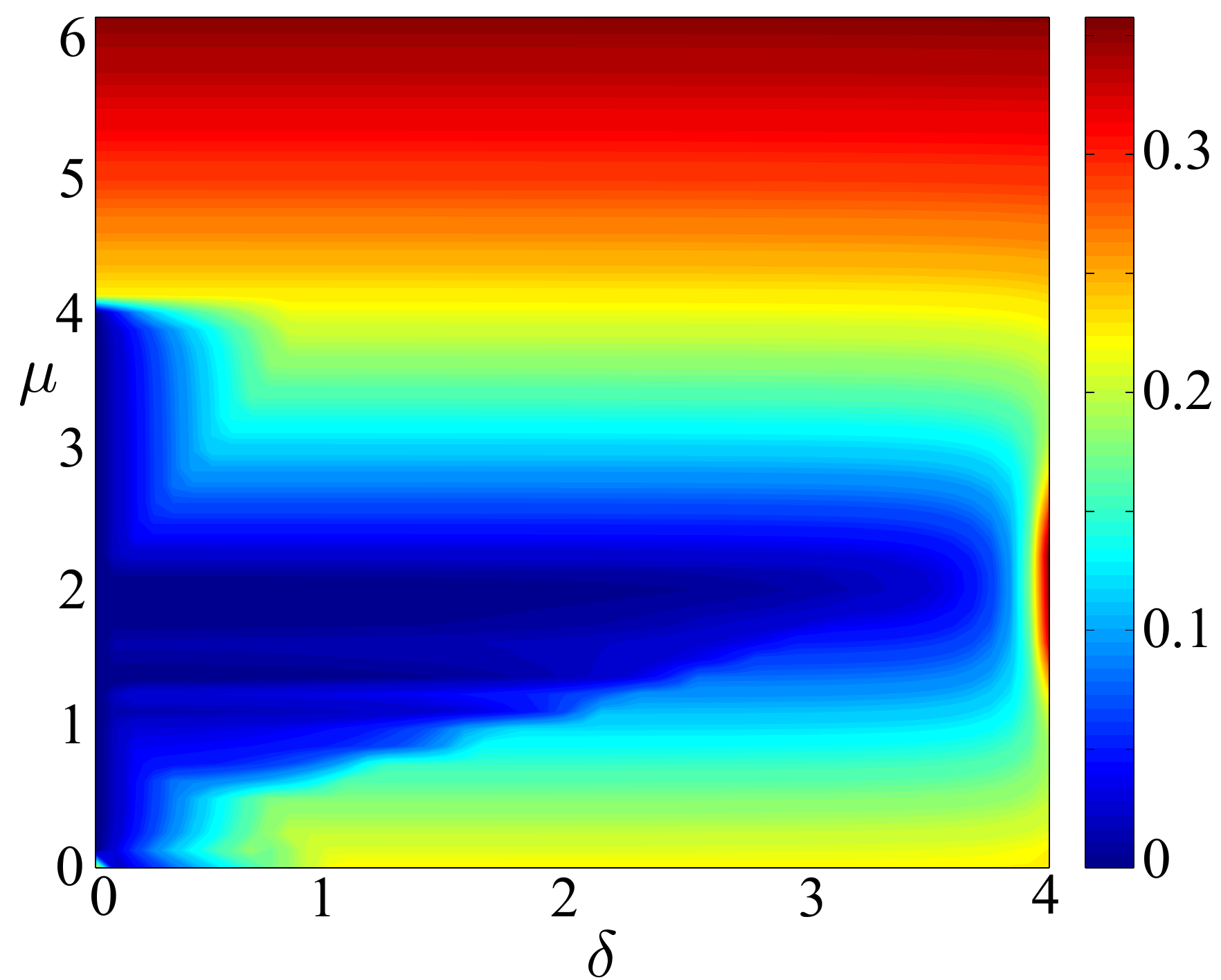}
\caption{ \label{PD_ent} The sublattice entanglement, as characterized by the minimum value $\min_{\bf p}|{\bf S}_{b/w}(\p)|$, as functions of $\mu$ and $\delta$. Comparison to Fig.~\ref{PD} shows that the winding number $\tilde{\nu}$ reproduces the Chern number everywhere except where the sublattices are close to being maximally entangled ($\min_{\bf p}|{\bf S}_{b/w}(\p)| \lesssim 0.1$). In these regimes numerical errors due the momentum space coarse graining become significant as $|{\bf S}_{b/w}(\p)|$ becomes very small.}  
\end{figure}



\subsubsection{Adiabatic connection to Kitaev's honeycomb model}

We show in Appendix \ref{App_honeycomb} that our model is adiabatically connected to Kitaev's honeycomb model \cite{Kitaev06}. This connection, which is schematically illustrated in Fig.~\ref{lattice},  enables us immediately to understand some of the features of the phase diagram of our model.

First of all,  in the limit $\mu \gg \delta$ the sign staggering becomes negligible, and when also $\mu \gg t$ the resulting $\nu=0$ phase should be identified with a strong pairing like phase. In the honeycomb model it corresponds to the dimerized phase, where the vortices, while exhibiting semionic statistics, do not bind Majorana modes. Here we are interested in the regime where the detuning $\delta$ is comparable to $\mu$. This regime supports topological phases characterized by Chern numbers $\nu=\pm 1$  and $\nu=\pm2$ phases, that emerge in the weakly ($\mu \gtrsim \delta/2$) and strongly ($\mu \lesssim \delta/2$) staggered regimes, respectively. The adiabatic connection to the hoenycomb model reveals that sufficiently staggered chemical potential is equivalent to the presence of a background vortex lattice. In particular, we find that the $\nu=-1$ phase in the weakly staggered regime corresponds to the absence of a lattice of $\pi$-flux vortices, while the $\nu=2$ phase in the strongly staggered regime corresponds to the presence of one \cite{Lahtinen10}.

The reason the presence of a vortex lattice in the honeycomb model gives rise to a Chern number $\nu=\pm2$ phase can be traced back to the properties of the localized Majorana modes present in the model. The Chern number $\nu=\pm 1$ phases in the weakly staggered regime are adiabatically connected to the non-Abelian phase of the honeycomb model, where the vortices have explicitly been shown to bind Majorana modes with short range interactions \cite{Lahtinen11}. By increasing the chemical potential staggering a lattice of these vortices is introduced. The interactions imply that the Majorana modes can hybridize and form a collective topological state. This mechanism of topological liquid nucleation has been studied in~\cite{Lahtinen12}, where one finds that for regular vortex lattices the resulting state is always of Abelian nature (characterized by an even Chern number). Importantly, this collective state is unique -- switching on the vortex lattice will only result in this state if the vortices bind Majorana modes. This implies that the nucleation mechanism could be used as an alternative global probe for the existence of Majorana modes in the model: Detection of the Chern number change as the vortex lattice is introduced (staggering is increased) would provide direct evidence for the existence of Majorana modes in the model.

\subsection{Detection of the phase diagram from the observables}

Fig.~\ref{PD} shows the comparison between the Chern numbers calculated from the ground state and the winding number \rf{winding} calculated from the observables \rf{eqn:manyobs} for the black and white sublattices. In general, we find excellent agreement between the two invariants. The only discrepancies occur in regions where the spectral gap is small. As anticipated in Section II.B, we can attribute this to the sublattices becoming close to maximally entanglement. Fig.~\ref{PD_ent} shows that in regimes where the norm $|\mathbf{S}_{b/w}|$ becomes small, thus causing numerical errors due to momentum space coarse graining. So only $|\nu|$ may be captured (which however is still sufficient to characterize the type of topological order). Everywhere else the full Chern number is accurately reproduced.  Thus the sublattice entanglement, as measured by the norm $|\mathbf{S}_{b/w}|$, indeed provides a good experimental measure for the fidelity of the winding number \rf{winding}.

\subsubsection{Distinguishing topological phases by only density measurements}

\begin{figure}
\includegraphics[width=8.5cm]{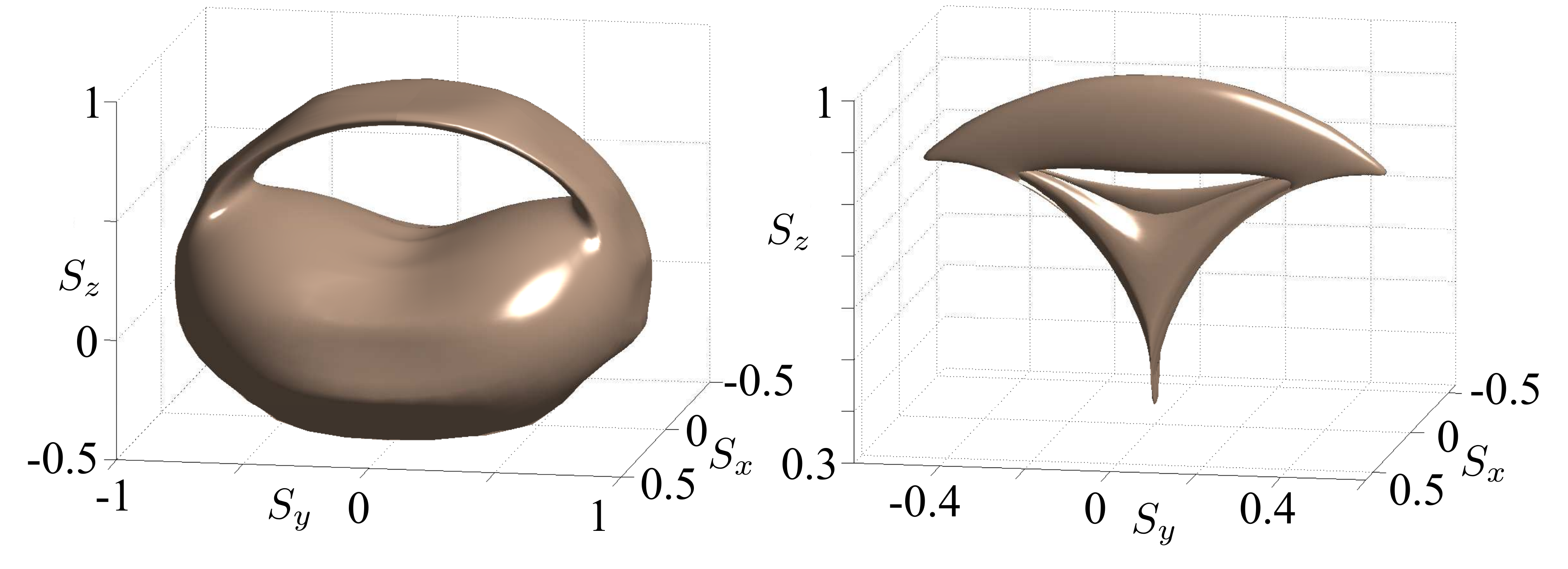}
\caption{Illustration of the vector fields ${\bf S}_{(i)}(\p)$ for a $\nu=-1$ phase. Here we plot the values of {\bf S}$_{(b)}$(left) and {\bf S}$_{(w)}$(right) for ($\delta,\mu$)=(1,3). It can be seen that {\bf S}$_{(b)}$ winds once around the origin, thus giving a partial $|\tilde{\nu}_b|=1$ contribution, while {\bf S}$_{(w)}$ does not enclose the origin so it gives zero contribution. Thus we verify that $|\nu|=|\tilde{\nu}_b|+|\tilde{\nu}_w|$} \label{fig:Doughnut}
\end{figure}

\begin{figure}[t]
\includegraphics[width=8.5cm]{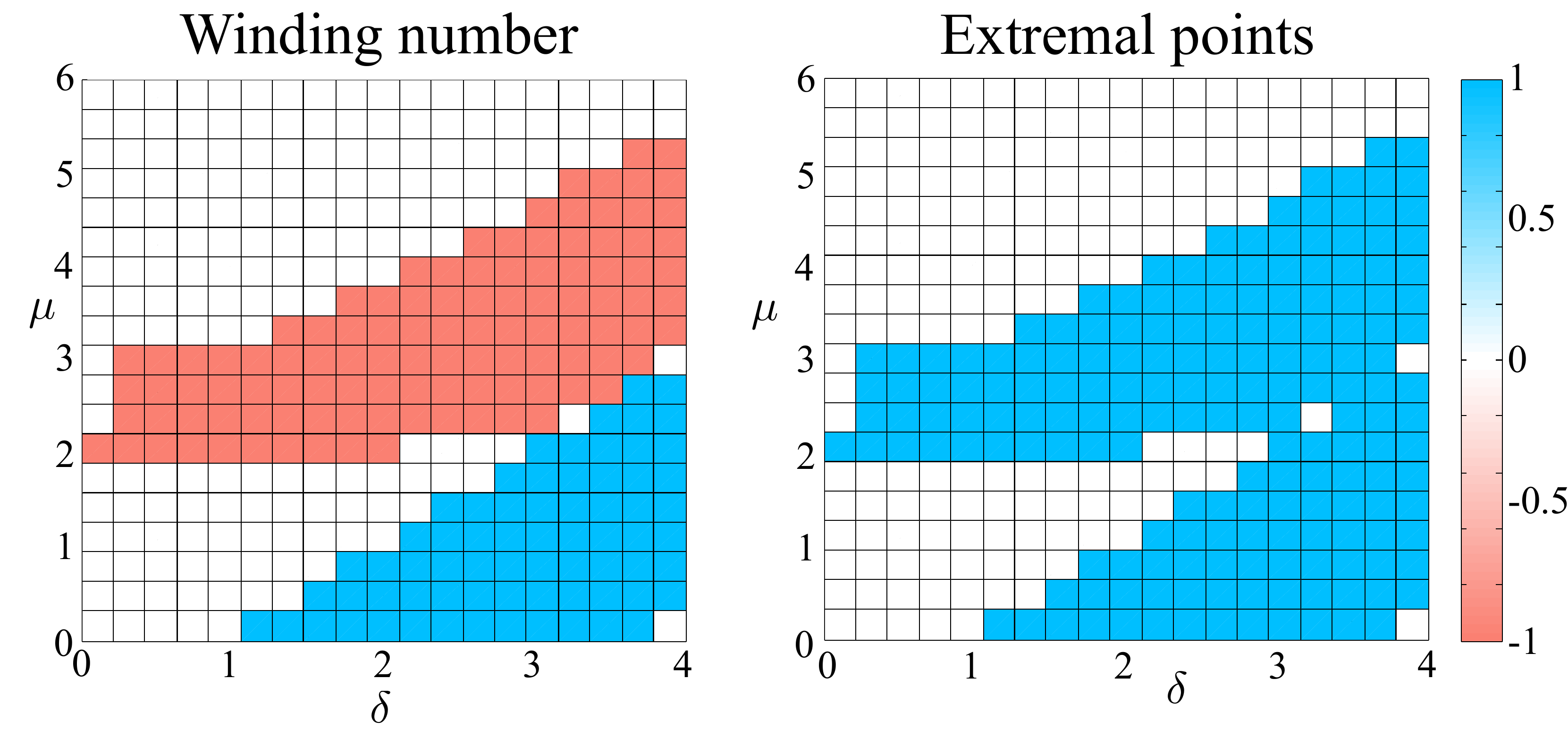}
\caption{Comparison between the winding number $\tilde{\nu}_{(w)}$ (left) and its parity as computed from the null-gradient-points of $S^z_{w}$ (right). The parity is in perfect agreement across the phase diagram. The simulation is performed on $20\times10$ lattice sites.  } \label{fig:Extremalpoints}
\end{figure}

In Section II.C we argued that the parity of the winding numbers should be detectable from the density measurements only. These correspond to $\Sigma^z_{(i)}$ measurements that, when applied to our staggered model, will give the compact surfaces $\mathbf{S}_b(\p)$ and $\mathbf{S}_w(\p)$ (see Fig.~\ref{fig:Doughnut} for an illustration).  By applying the protocol of counting the saddle points, assigning the parities $|\tilde{\nu}_b|$ and $|\tilde{\nu}_w|$ accordingly and adding them up, Fig.~\ref{fig:Extremalpoints} shows that we can accurately reproduce the absolute value of the Chern number everywhere in the phase diagram. To be precise, we find that the following always holds:   (i) $N=|\tilde\nu_{b}|+|\tilde\nu_w|=0$ coincides always with the trivial $\nu=0$ phase, (ii) $N=1$ corresponds always to the non-Abelian topological phase with $|\nu|=1$, and (iii) we find $N=2$ only when the system is in the $|\nu|=2$ phase.  Thus the experimentally accessible density measurements are sufficient to distinguish between all the topological phases of our model.

\subsubsection{Robustness to perturbations} 

So far we shown that our detection scheme based on decomposition to sublattice observables accurately captures the phase diagram of our model except for special regions where the sublattices are too entangled. While this imposes accuracy limitations when applying the scheme, one may also ask how reliable the scheme is to the presence of perturbations in the Hamiltonian \rf{H2}. In Section IV we propose an optical lattice implementation of our model. Here, we consider two general types of imperfections that one expects to be present in cold atom experiments: a harmonic trapping potential that breaks translational invariance, and finite temperature.

\begin{figure}
\includegraphics[width=\linewidth]{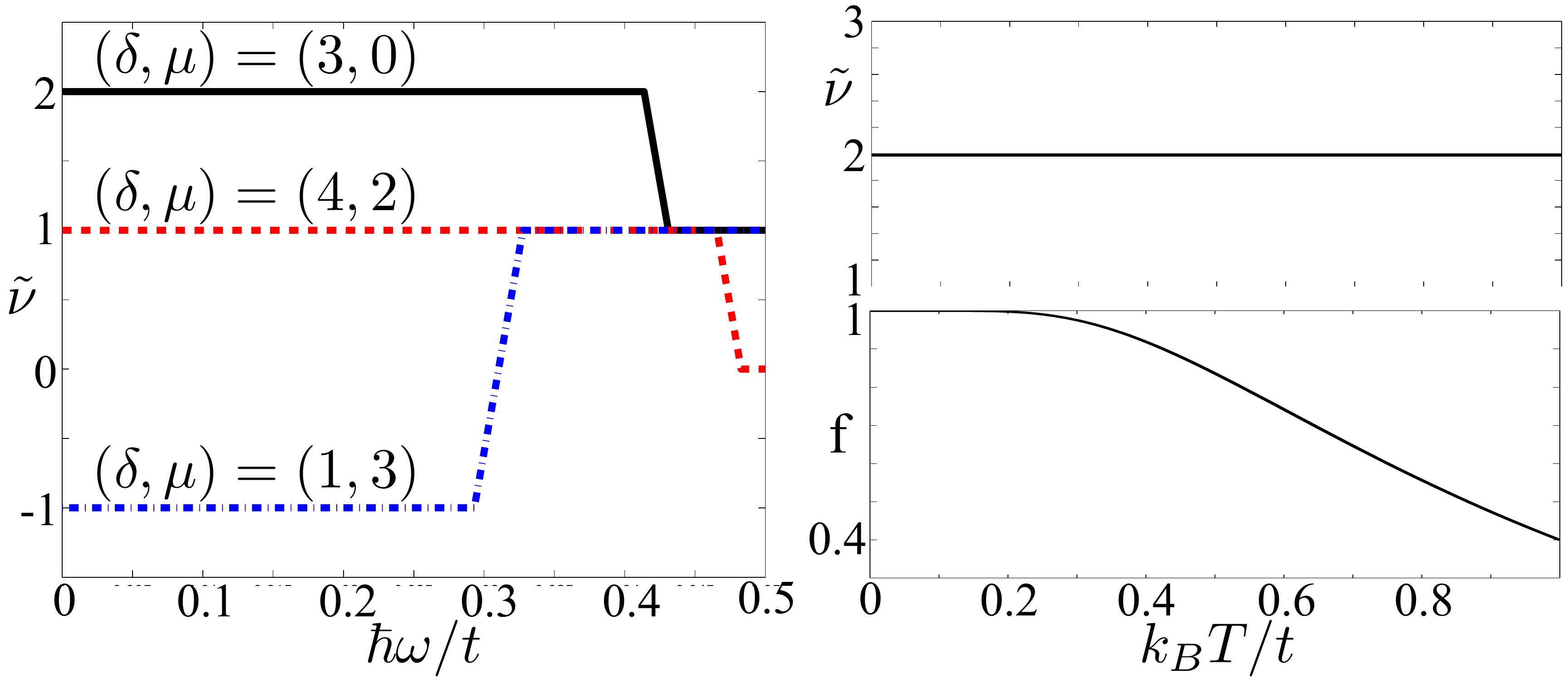}
\caption{\label{Resproblem} \textit{Left:} Winding numbers $\tilde{\nu}$ as functions of the trapping frequency $\omega$ in a finite $15\times15$ site system. \textit{Right:} The winding number and the visibility, i.e. norm of the vector field $\textbf{S}^{\mathrm{th}}_{(i)}(\textbf{p},T)=f(k_B T) \textbf{S}_{(i)}(\textbf{p},0)$ at finite temperature $T$, in a uniform system without a trap.}
\end{figure}

We simulate the trap in a finite $L \times L$ lattice with open boundary conditions by introducing in \rf{H2} the chemical potential $\mu_{\bf j} = \mu+(-1)^{j_x}\delta+M d^2 \omega^2[(j_x-L/2)^2+(j_y-L/2)^2]$, where $M$ is the mass of the atomic species and $d$ is the lattice spacing. Assuming that a local density approximation holds~\cite{Jaksch98}, a spatially dependent chemical potential induces in general the coexistence of different phases: some of insulating character, some not; some with topological order, some with no order at all. The Chern number is no longer defined in the absence of translational invariance. However, the winding number~(\ref{Chern}) can still be used to identify the existence of topological order, because regions in a trivial phase do not contribute to the expectation values $\textbf{S}_{(i)}$~\cite{Alba11}. Indeed, Fig.~\ref{Resproblem} shows that all topological phases are robust for a wide range of trapping frequencies $\omega$. We conclude that at least for small perturbing potentials the winding number \rf{winding} will still offer a reliable characterization of the phase diagram. 

To model the effect of finite temperature $T$ we restrict to fermionic excitations in the lower band with no thermal vortex excitations. The thermal state is then a product state in the momentum space. Computing the  expectation values \rf{eqn:Sp} both numerically and analytically, we find that temperature only leads to a change in the norm of the observables, ${\bf S}_{(i)}^{\mathrm{th}}(\mathbf{p},T) = f(k_B T) {\bf S}_{(i)}(\mathbf{p})$. While theoretically such effect can just be normalized away, experimentally this corresponds to a reduced visibility ($0<f(k_BT)<1$) in the time-of-flight measurements. Since the supression of the norm, exactly like high entanglement between subattices, makes it harder to obtain ${\bf S}_{(i)}(\mathbf{p})$ accurately, finite temperature implies that higher resolution measurements are required. Assuming that this is within the state-of-the-art experimental precision, we numerically verify in Fig.~\ref{Resproblem} that the winding number \rf{Chern} still faithfully reproduced. Thus we conclude that finite temperature can be compensated for by increased precision and therefore it does not pose a fundamental challenge for our detection scheme.

\section{Optical lattice implementation and the experimental detection of the winding numbers}

In this last section we first outline a scheme to implement our staggered model with cold atoms in an optical lattice. We then show how to recover, from time of flight images in this particular setup, the winding numbers with which the phase diagram from Fig.~\ref{PD}b can be experimentally reconstructed. A quantitative analysis of the parameters for a particular implementation is left for Appendix \ref{App_implementation}.

\subsection{Optical lattice implementation} 

As Hamiltonian~\rf{H2} describes spinless fermions, it can be implemented with atoms in a single internal state only. However, it can also be implemented with two atomic states, which can be advantageous for two reasons. First, by trapping the distinct atomic states in a checkerboard state-dependent optical lattice, denoted by the circles and squares in Fig.~\ref{lattice}, we can use Raman-assisted tunnelling~\cite{Jaksch03, Gerbier10, Mazza12} to implement both the complex tunnelling amplitudes and control the chemical potentials. Second, using two atomic states we can implement the pairing terms between neighbouring sites using $s$-wave Feshbach resonances~\cite{Bruun04, AllStars}. If we were using only a single atomic state, the Pauli exclusion principle would force us to employ $p$-wave Feshbach resonances, which are harder to observe~\cite{Regal03,Guenter05}.

We propose to generate the lattice of model \rf{H2} by focusing the diffracted image from a holographic mask onto the focal plane of an extremely large aperture lens~\cite{Greiner09}. The sublattices host different hyperfine states of the same atomic species which are sensitive to different polarisations of the trapping laser beams. Both sublattices can be displaced one on top of another by means of an electro-optic modulator~\cite{Chin10}.

\begin{figure}
\includegraphics[width=8cm]{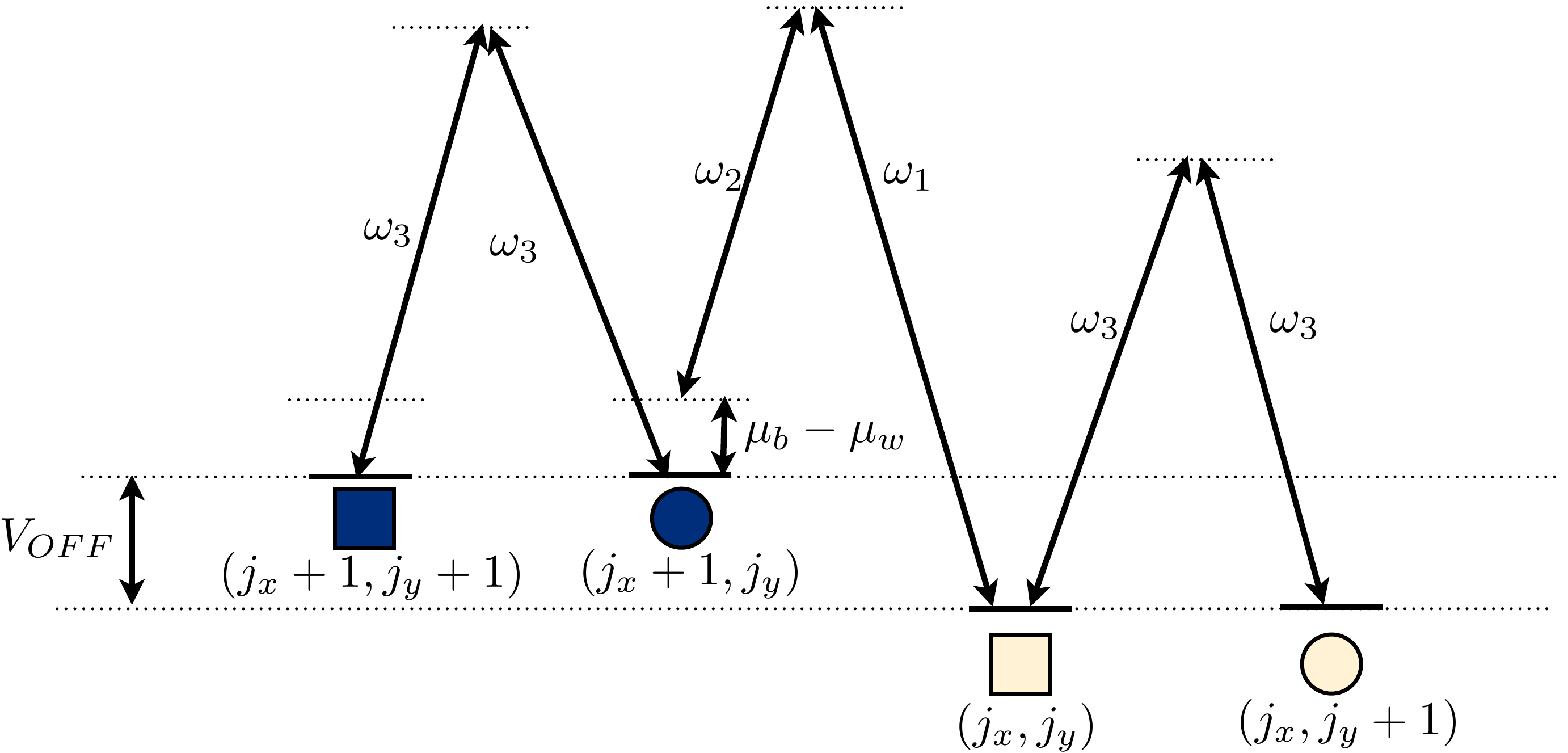}
\caption{Energy level structure, showing how two sets of Raman beams
  allow for independent transitions due to the presence of the offset
  $V_{OFF}$. This offset can be tuned to allow for different values of the difference in chemical potentials $\mu_w-\mu_b$. The indices $(j_x,j_y)$ stand for horizontal and vertical
  position on the lattice, respectively.} \label{fig:04}
\end{figure}

We also superimpose a superlattice potential $V=V_{OFF} \sin^{2}(\pi x/\sqrt{2} d)$, where $d$ is the lattice constant for each sublattice. $V_{OFF}$ can be controlled by changing the intensity of the laser beam, which creates this potential. This superlattice structure effectively adds a tuneable energy offset $V_{OFF}$ to every other column in the lattice. This offset serves a two-fold purpose: it allows for independent Raman tunnelling in each direction and acts as a knob for changing the value of the chemical potential difference $\mu_w-\mu_b$, as it is shown in Fig.~\ref{fig:04}.

Our implementation employs a Raman-assisted tunnelling scheme on an optical lattice with a pattern of phases~\cite{Jaksch03, Gerbier10,Mazza12}, as shown in Fig.~\ref{mapping}. In this figure we use a convention that the tunnelling direction is set to go from one species (circle) to the other (square). Reversing the direction complex conjugates the tunnelling amplitude. The tunnelling element between sites ${\bf j}$ and ${\bf j}'$ in a Raman transition assisted by two lasers of wavevectors ${\bf k}_1$ and ${\bf k}_2$ of amplitude $\Omega_R$ can be parameterised as $t= e^{i {\bf q} \cdot {\bf r}_+} t_0(d,{\bf q})$, where $t_0$ is a real number which only depends on the nearest-neighbour distance $d$, the Raman frequency $\Omega_R$ and the difference between the Raman beam wavevectors ${\bf q}={\bf k}_1-{\bf k}_2$. Also ${\bf r}_+=({\bf j}+{\bf j}')/2$ is the midpoint between the two neighbouring sites. The phase of the hopping parameter is thus determined by wavevector ${\bf q}$. We can see in Fig.~\ref{lattice} that two different phase wavelengths for the horizontal and vertical transition amplitudes are needed. So implementation of this hopping pattern requires a Raman pair for each direction. Fig.~\ref{fig:04} shows the energy level structure which accomplishes this pattern.

In order to study the experimental feasibility of the proposal, we have considered a possible implementation with $^{40}$K atoms in a state-dependent optical lattice (see Appendix \ref{App_implementation}). We expect a small heating rate~\cite{Alba10} of about 1 Hz which is the dominant time-scale for decoherence. Using a potential depth of about 22 recoil energies, ordinary hopping can be suppressed in each sublattice, while still having a significant overlap between neighbouring wave functions ---the quantity that determines the strength of both $t$ and $\Delta$. Typical estimates for the Raman-assisted tunnelling~\cite{Jaksch03, Gerbier10, Mazza12} and the induced $s$-wave pairing~\cite{AllStars}, give us an estimate of about 1 kHz for $t$, and $0.5$ kHz for $\Delta$. These numbers could be improved through the use of alkaline-earth atoms~\cite{Gerbier10}.

\subsection{Experimental construction of the winding numbers from time-of-flight images}

To obtain the full winding number $\tilde \nu$ one need to construct the independent integer-valued winding number $\tilde{\nu}_{(i)}$ for each pseudospin component. In the case of our model,  the psedospin components coincide with the ``black'' ($i=b$) and ``white'' ($i=w$) sublattices that are distinguished by their different chemical potential. In an optical lattice implementation, this energy offset between the atoms can be employed to release them from one of the two sublattices at a time and thus the observables $\SSIGMA_{(i),\p}$ for each sublattice can be independently evaluated. We outline below a general protocol to obtain all components of the vectors $\mathbf{S}_{(i)}(\p)$ from which the sublattice winding numbers $\tilde{\nu}_{(i)}$ can be constructed.

The experimental measurement of the operators \rf{eqn:manyobs} in an optical lattice setting employs the fact that time-of-flight images give direct access to the momentum space densities $\langle n_{(i),\bf p}\rangle = \langle a^\dagger_{(i),\bf p} a_{(i),\bf p}\rangle$. These are sufficient to fully determine $S^z_{(i)}$, which, as discussed in Section III.B, enables to unambiguously distinguish between all the distinct types topological phases (different $|\nu|$). Thus the time-of-flight images, a standard technique in optical lattice experiments, are sufficient to identify the phases of our model.

In order to construct the full winding number one needs to obtain also the orthogonal components $S^x_{(i)}$ and $S^y_{(i)}$. They can in general be obtained by suitably switching off the pairing and tunnelling terms of \rf{H2} before releasing the atoms from the trap. This will rotate the observables $\Sigma^{x,y}_{(i),\p}$ to $\Sigma^{z}_{(i),\p}$, which can then be measured from time-of-flight images as above. For instance, when hopping in both directions and pairing in $x$-direction is suppressed, e.g. by raising the lattice in this direction, the Hamiltonian \rf{H2} acquires the form
\begin{equation}
H_\text{rot}=\Delta\sin(p_y)i\left(a^\dagger_{(i),\bf p}a^\dagger_{(i),\bf p}-\mathrm{H.c.}\right)\propto \Sigma^{y}_{(i),\bf p}.
\end{equation}
This operator implements a rotation around the $S^y$ axis, mapping the value of the $S^x$ operator onto the $S^z$ axis, which after time $t$ gives
\begin{equation} \label{evol}
  S^z(\p,t) = \cos(\theta_\p)S^z(\p,0) + \sin(\theta_\p)S^x(\p,0),
\end{equation}
with $\theta=\Delta\sin(p_y)t/\hbar$. Time of flight image can again be used to measure this quantity from which the value of $S^x$ can be extracted once the unrotated component $S^z(\p,0)$ has been determined. Finally, the value of $S^y$ can be obtained experimentally using a similar two-step process as above. Evolving the system with only hopping along the $y$ direction maps $S^y$ to $S^x$, which when followed by a pairing evolution can again be mapped to the directly observable $S^z$.

The dependence of the evolution \rf{evol} on the momentum $p_y$ implies that the Hamiltonian rotations around $p_y = 0, \pm\pi/2, \pm\pi$ will be infinitely slow. This experimental challenge can be overcome in two ways. One way is to numerically post-process the measurements by extrapolating smoothly the values of $\bf S$ from the measurements of $S^z,S^x$ and $S^y$. We have numerically verified that given $|\bf S|$ does not become zero anywhere, and that the angles in the $xy$ plane behave smoothly across the Brillouin zone, this can be efficiently performed. An alternative is to use additional complementary noise correlation measurements $\langle n_{(i),\bf p} n_{(i),-\bf p}\rangle$. Using Wick's theorem such an observable can be written in the form
\begin{align}
\langle n_{(i),\bf p} n_{(i),-\bf p}\rangle &= \langle n_{(i),\bf p}\rangle\langle n_{(i),-\bf p}\rangle +
|\langle a^\dagger_{(i),\bf p}a^\dagger_{(i),-\bf p}\rangle|^2 \\
&+ \langle a^\dagger_{(i),\bf p} a_{(i),-\bf p}\rangle
\langle a_{(i),\bf p} a^\dagger_{(i),-\bf p}\rangle. \nonumber
\end{align}
As $\langle n_{(i),\bf p}\rangle$ follows from the usual time of flight images and $\langle a_{(i),\bf p} a^\dagger_{(i),-\bf p}\rangle$ can be obtained from them after Bragg scattering with momentum $2\bf p$ (for our model they always vanish), in essence noise correlations give us access to the orthogonal projection of the pseudo-spin components, $(S^\perp)^2=(S^x)^2 + (S^y)^2$. Thus once $S^{x}$ has been obtained, the noise correlations can be employed as an alternative way to obtain $S^{y}$.

\section{Conclusions}

We have presented a general method to detect the Chern number of superconducting models from time-of-flight images. This method is readily applicable to any topological superconducting state regardless of the microscopic realization \cite{Sato09,Zhang08,AllStars}. The only requirement is the ability to measure independently the relevant operators for each pseudospin component, such as spin orientation, internal atomic states or sublattices due to staggering. While not restricted only to, our method is particularly suited for optical lattice experiments where time-of-flight images, a standard technique, readily give access to the relevant operators. We presented a full set of experimental manipulations for the reconstruction of the Chern number. We also showed that the time-of-flight images without additional manipulation can give sufficient information (the absolute value of the Chern number) to distinguish between the different types of topological order. With the preparation of topologically ordered states with cold atoms in optical lattices as the ultimate goal, this provides a simple and reliable diagnostic tool to probe the nature of the prepared states.

To demonstrate our detection scheme, we applied it to a model of staggered spinless fermions with $s$-wave pairing, a new route to topological phases with cold atoms. We could robustly identify topological phases with Chern numbers $\nu=0$, $\nu= \pm 1$ and $\nu=\pm2$. The few disagreeing parameters regimes were found to correlate with high sublattice entanglement, which in itself is a physical observable. Thus the detection scheme has an in-built fidelity measure that can be used to evaluate its reliability in reproducing the Chern numbers. Furthermore, we showed that the detection scheme remains robust under two omnipresent perturbations in cold atom experiments: translational invariance breaking trapping potential and finite temperature. The latter could be compensated for by increasing detection precision, which contrasts with the behaviour of topological entanglement entropy, an alternative probe for topological order in cold atom systems~\cite{Abanin12}. In the thermodynamic limit it vanishes at any finite temperature rendering its applicability challenging~\cite{Castelnovo07,Iblisdir09}. In addition, unlike our method topological entropy can not distinguish topological phases with same total quantum dimensions~\cite{Pachos12}.

Finally, we explicitly demonstrated that the proposed model of staggered spinless fermions with $s$-wave pairing could be adiabatically connected to Kitaev's honeycomb model \cite{Kitaev06}. The proposed optical lattice implementation would thus offer an alternative route for realizing this celebrated model. In our realization we could relate the staggering in the chemical potential, an experimentally accessible parameter, to the presence or absence of a background vortex lattice. We showed that the presence of such a lattice underlies the Chern number $\nu=\pm 2$ phases, and that these phases should be understood as a unique collective states of Majorana modes bound to the vortices, as studied in detail in \cite{Lahtinen12}. As this phase can only arise as the collective state of Majorana modes, detecting the change in the Chern number when the vortex lattice is switched on provides a global probe for the existence of Majorana modes.


\acknowledgements

JKP would like to thank Wolfgang Ketterle for inspiring conversations. This work was supported by EPSRC and by Spanish MICINN Project FIS2009-10061, Beca FPU No. AP 2009-1761, CAM research consortium QUITEMAD S2009-ESP-1594.

\appendix
\section{Chiral topological order with $s$-wave pairing}
\label{App_model}

In this Appendix we first give the analytic solution to our staggered superconducting model. Then we verify the existence of edge states that together with particle-hole symmetry imply that the phases with odd Chern numbers support localized Majorana modes. Finally, we discuss the interpretation of the staggered tunnelling as an effective spin-orbit coupling.

\subsection{Analytic solution}

The Hamiltonian \rf{H2} can be Fourier transformed with respect to the two site unit cell illustrated in Fig.~\ref{lattice}. Writing it in the particle-hole basis $\PSI^\dagger_\p=(a^\dagger_{b,\p}, a^\dagger_{w,\p},a_{b,-\p}, a_{w,-\p})$, we obtain the quadratic Bogoliubov-de Gennes Hamiltonian $H = \int_{BZ} \PSI^\dagger_\p H(\p) \PSI_\p \mathrm{d}^2p$, where
\be \label{H_Bloch}
H(\mathbf{p}) = \left( \begin{array}{cccc} f_+ & ig_+^* & ih & g_-^* \\
 -ig_+ & f_- & -g_- & ih \\
 -ih & -g_-^* & -f_+ & ig_+^* \\
 g_- & -ih & -ig_+ & -f_- \end{array} \right),
\ee
with 
\be
\begin{array}{rcl}
 	f_\pm & = & (\mu \pm \delta) + 2t\cos(p_y), \\
 	g_+ & = & t(1 + e^{2ip_x}), \\
 	g_- & = & \Delta(1 - e^{2ip_x}), \\
 	h & = & 2\Delta \sin(p_y).
 	\end{array} \nonumber
\ee
The Hamiltonian can be diagonalised with a Bogoliubov transformation, which gives the four particle-hole symmetric energy bands
\be \label{bands}
	E_n^\pm(\p) = \pm \sqrt{A(\p)+(-1)^n\sqrt{A^2(\p)-4B(\p)}},
\ee
where
\be
\begin{array}{rcl} \nonumber
	A(\p) & = & f_+^2+f_-^2+4\left( |g_+|^2 + h^2 + |g_-|^2 \right), \\
	B(\p) & = & |g_+|^4+h^4+|g_-|^4 + f_+^2 f_-^2 + \\
	\ & \ & h^2 (f_+^2+f_-^2) - 2f_+f_-(|g_+|^2-|g_-|^2) - \\ 
	\ & \ & 2h^2(|g_+|^2+|g_-|^2)-2\textrm{Re}(g_- g_+^*)^2.
\end{array}
\ee

\begin{figure}[t]
\includegraphics[width=0.48\textwidth]{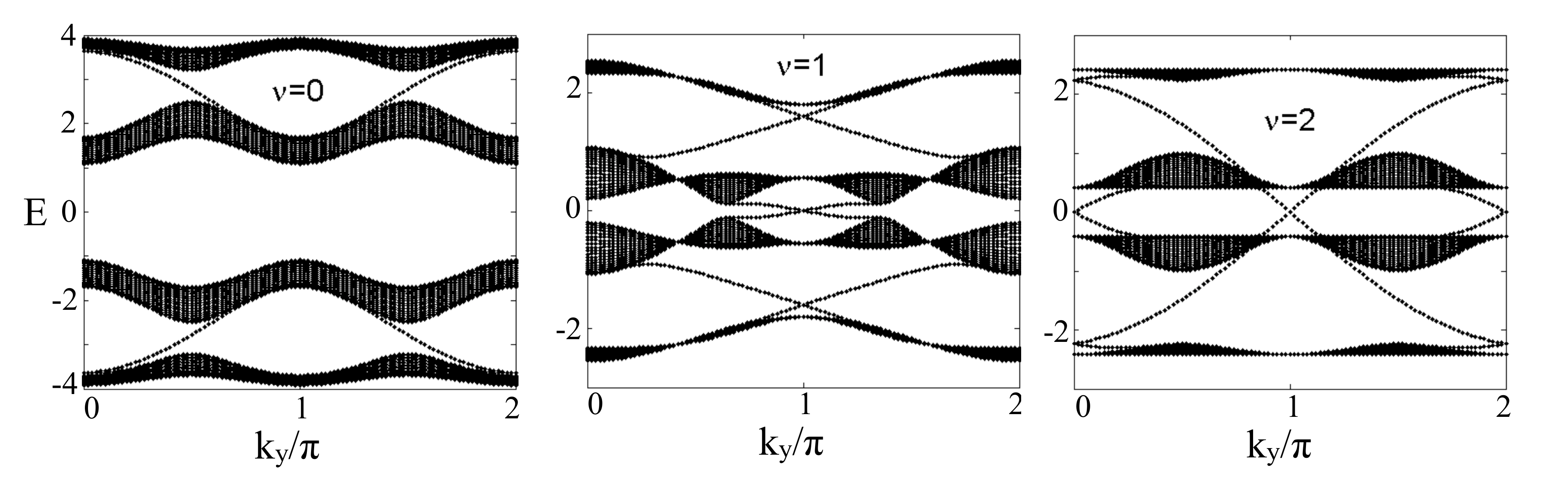} 
\caption{\label{edgestates} Edge states in the different topological phases. The spectral flow on a cylinder (open boundary conditions in $x$-direction) for the (a) $\nu=0$ [$(\delta,\mu)=(5,0)$], (b) $\nu=1$ [$(\delta,\mu)=(5,2)$] and (c) $\nu=2$ [$(\delta,\mu) = (2,0)$] phases shows $|\nu|$ edge states per edge crossing zero energy. The plots are for $t=\Delta=1$.}
\end{figure}

The particle-hole symmetry is represented by $C=\sigma^x \otimes \1$ that swaps the creation and annihilation operators of opposite momenta. It acts on \rf{H_Bloch} as 
\be \label{PH}
CH(\p)C^{-1}=-H^*(-\p),
\ee
which implies that zero energy eigenstates at the momenta $\p=(0,0),(0,\pi)$ will be self-conjugate. Fig.~\ref{edgestates} shows that in the $\nu=1$ and $\nu=2$ phases the edge states indeed cross zero energy at these momenta, implying that they are (dispersing) Majorana modes. In the presence of a vortex (a puncture in the plane with $\pi$-flux through it), they will thus become localised at the vortex cores \cite{Chamon1}. Odd number of edge states (odd $\nu$) implies that an isolated Majorana mode will always remain localised at zero energy at the vortex core, while an even number of them (even $\nu$) leads to complete hybridisation with all the Majorana modes pairing up to localised Dirac fermions. In the $\nu=0$ phases no edge states cross zero energy (although high energy edge states can still exist as shown in Fig.~\ref{edgestates}), and vortices will not bind localised low-energy states of either Majorana or Dirac type. 

\subsection{Staggered tunnelling as an effective spin-orbit coupling}

Another way of understanding the emergence of localized Majorana modes is to consider our model as an anisotropic analogue of spin-orbit coupled systems in the proximity of a regular $s$-wave superconductor \cite{Alicea10}. Let us consider the different couplings of the Hamiltonian \rf{H2} separately.

Staggering in the  tunnelling phases and in chemical potential breaks translational symmetry to a subgroup such that the system is still translationally invariant with respect to a two site ``magnetic" unit cell. For the coupling pattern shown in Fig. \ref{lattice}, we colour these two sites as black ($b$) and white ($w$). This sublattice degree of freedom can be interpreted as a pseudospin $\tau \in (b,w)$ of the fermions $a_{{\bf j},\tau}^\dagger$. Using the ``spinor'' $\psi_{\bf j}^\dagger=(a_{b,{\bf j}}^\dagger,a_{w,{\bf j}}^\dagger)$, we can rewrite the different terms of \rf{H2} in the following way:
\bq \label{SOinterpret}
 \mu_{\bf j} a_{\bf j}^\dagger a_{{\bf j}} & \to & \mu \psi_{\bf j}^\dagger \psi_{{\bf j}} + V_z \psi_{\bf j}^\dagger \tau^z \psi_{{\bf j}}, \nonumber \\
 i(-1)^{j_x}ta_{\bf j}^\dagger a_{{\bf j}+\hat{\bf x}} & \to & \alpha \psi_{\bf j}^\dagger \tau^y \psi_{{\bf j}+\hat{\bf x}}+V_y \psi_{\bf j}^\dagger \tau^y \psi_{{\bf j}}, \nonumber \\
 ta_{\bf j}^\dagger a_{{\bf j}+\hat{\bf y}} & \to & t \psi_{\bf j}^\dagger \psi_{{\bf j}+\hat{\bf y}}, \\
 \Delta a_{\bf j}^\dagger a_{{\bf j}+\hat{\bf x}}^\dagger & \to & \Delta \psi_{\bf j}^\dagger \tau^x \psi_{{\bf j}}^\dagger + \Delta \psi_{\bf j}^\dagger \tau^x \psi_{{\bf j}+\hat{\bf x}}^\dagger, \nonumber \\
 \Delta a_{\bf j}^\dagger a_{{\bf j}+\hat{\bf y}}^\dagger & \to & \Delta \psi_{\bf j}^\dagger \psi_{{\bf j}+\hat{\bf y}}^\dagger. \nonumber
\eq
The Pauli matrices $\tau^\alpha$ act on the pseudospin degree of freedom. This suggests the following interpretation in terms of the fermions $\psi_{\bf j}^\dagger$:
\begin{itemize}
\item $\mu$ still acts as the chemical potential, while the detuning acts now effectively as a Zeeman term of magnitude $V_z=\delta$.
\item  Tunnelling in $x$-direction realises an anisotropic Rashba type spin-orbit coupling $p_x \sigma^y$ of magnitude $\alpha=t$ and a transverse magnetic field of magnitude $V_y=t$.
\item Pairing will be of uniform amplitude $\Delta$, but it will be an anisotropic mixture of singlet ($x$-direction) and triplet pairing ($y$-direction).
\end{itemize}
The elements above -- the spin-orbit coupling, magnetic fields of different direction and the $s$-wave pairing -- are the components of Majorana mode hosting semiconductor heterostructures \cite{Alicea10}. It would be interesting to study how far the analogy between staggered tunneling and spin-orbit coupling could be pushed.


\section{Adiabatic connection to Kitaev's honeycomb lattice model}
\label{App_honeycomb}

In this section we demonstrate that our staggered model \rf{H2} is adiabatically connected to Kitaev's honeycomb model \cite{Kitaev06}, which is known to support localised Majorana modes with short-range interactions \cite{Lahtinen11}. We show this explicitly for the $\nu=2$ phase, which we connect to the $\nu=2$ phase arising in the full-vortex sector as the unique collective state of the Majorana modes bound at the vortex cores \cite{Lahtinen10, Lahtinen12}.

In nutshell, the honeycomb model is a local spin lattice model that contains nearest-neighbour two-spin interactions (of magnitudes $J_x, J_y$ and $J_z$ depending on link orientations) and next-nearest-neighbour three-spin interactions (of magnitude $K$) that break time-reversal symmetry. When mapped to a tight-binding model of free Majorana fermions on the honeycomb lattice, the spin interactions map into nearest and next nearest neighbour tunnelling, respectively. The model becomes exactly solvable when restricted to a particular symmetry sector that corresponds to some background pattern of $\pi$-flux vortices. \cite{Kitaev06}

\begin{figure}[t]
\includegraphics[width=8.2cm]{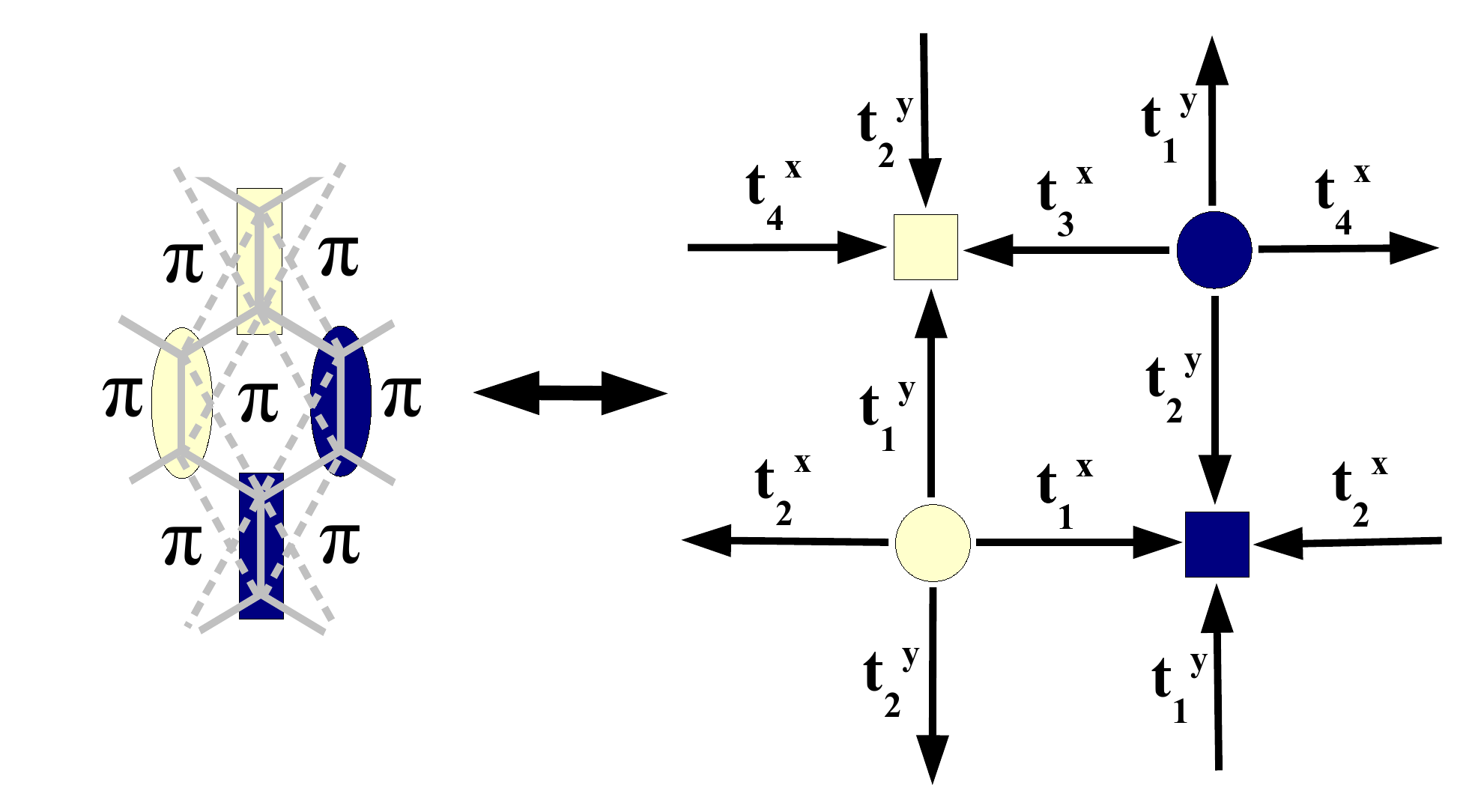}
\caption{ {\it Left:} The honeycomb model \rf{Hfv} with a $\pi$-flux on every plaquette. When mapped into a tight-binding model of Majorana fermions, the nearest neighbour hopping is along the solid links (of magnitude $J_z$ along the vertical links and $J_x$ or $J_y$ along the other two oriented links) and the included next-nearest-neighbour hopping along the dashed ones (of uniform magnitude $K$). The vortex lattice is encoded in the staggered signs of the $J_z$ hopping\cite{Lahtinen10}. When the Majorana fermions are paired into complex fermions, the vertical links become the sites of a square lattice, with $J_z$ translating into sign staggered chemical potential. {\it Right:} When the tunnelling couplings $t_{\bf j}^x$ and $t_{\bf j}^y$ in \rf{honey_hopping} are explicitly written out, one finds six independent couplings, which we denote as $t_1^x,t_2^x,t_3^x,t_4^x,t_1^y$ and $t_2^y$. Redefining the operators on the circle (square) sites by $c_{\bf j} \to e^{i\chi}c_{\bf j} (c_{\bf j} \to e^{-i\chi}c_{\bf j})$ preserves the real pairing potential for arbitrary $\chi$, while unitarily transforming the hopping amplitudes. For $\chi=\phi/2-\pi/4$ they are brought to the form \rf{coupling_deform}.   }\label{mapping}
\end{figure}

We are interested in the full-vortex sector ($\pi$-flux on each hexagonal plaquette), which supports topological phases with Chern numbers $\nu=0,\pm1$ and $\pm2$ \cite{Lahtinen10}. When the honeycomb model is restricted to it, the tight binding Hamiltonian can be written as \cite{Lahtinen08}
\bq \label{Hfv}
H_\mathrm{f.v.} \!\!\!&= & i \sum_{\bf j} \big[(-1)^{j_x} J_z a_{\bf j}b_{\bf j} +J_x a_{\bf j}b_{{\bf j}+\hat{\bf x}}+J_y a_{\bf j}b_{{\bf j}+\hat{\bf y}}\big] \\
\ & \ & \!\!\!\!\!\!\!\!\!\!+ iK \sum_{\bf j} (-1)^{j_x}\big[a_{\bf j} a_{{\bf j}-\hat{\bf x}} +a_{\bf j} a_{{\bf j}+\hat{\bf y}} + b_{\bf j} b_{{\bf j}+\hat{\bf x}} + b_{\bf j} b_{{\bf j}-\hat{\bf y}} \big], \nonumber
\eq
where $a^\dagger_{\bf j}=a_{\bf j}$ and $b^\dagger_{\bf j}=b_{\bf j}$ are Majorana operators on the two triangular sublattices of the honeycomb lattice. To simplify the demonstration of the adiabatic connection, we have included only four out of the six possible next nearest neighbour hoppings, as illustrated in Fig.~\ref{mapping}. The omitted terms are $a_{\bf j} a_{{\bf j}+{\bf \hat{x}+\hat{y}}}$ and $b_{\bf j} b_{{\bf j}+{\bf \hat{x}+\hat{y}}}$, that have been shown to be adiabatically tuneable to zero while staying in the same phase \cite{Yu08}.  

The full-vortex sector is encoded in the $(-1)^{j_x}$ factors that stagger the signs of the Majorana hopping amplitudes $J_z$ and $K$. Pairing the Majorana operators into complex fermions $c_{\bf j}$ by 
\be \label{Maj_pair}
a_{\bf j} = e^{i\theta_{\bf j}} c_{\bf j} + e^{-i\theta_{\bf j}} c_{\bf j}^\dagger, \qquad b_{\bf j} ={1\over i} (e^{i\theta_{\bf j}} c_{\bf j} - e^{-i\theta_{\bf j}} c_{\bf j}^\dagger),
\ee
the phase $\theta_{\bf j}$ to be defined below, the vertical links with couplings $J_z$ connecting the $a_{\bf j}$ and $b_{\bf j}$ sites of the honeycomb lattice become the sites of a square lattice, as illustrated in Fig.~\ref{mapping}. The Hamiltonian takes the form
\bq \label{H_honey_BCS}
H_\mathrm{f.v.} &=& \sum_{\bf j} \big[ \mu_{\bf j} c_{\bf j}^\dagger c_{\bf j} + t_{\bf j}^x c_{\bf j}^\dagger c_{{\bf j}+\hat{\bf x}} + t_{\bf j}^y c_{\bf j}^\dagger c_{{\bf j}+\hat{\bf y}}\no
& \ & \qquad \Delta_x c_{\bf j} c_{{\bf j}+\hat{\bf x}} + \Delta_y c_{\bf j} c_{{\bf j}+\hat{\bf y}}\big] + \mathrm{H.c.},
\eq
where we have defined
\be \label{honey_hopping}
\begin{array}{rcl}
\mu_{\bf j} & = & 2J_z(-1)^{j_x} \\
t_{\bf j}^x & = & r e^{i(-1)^{j_x} (2-(-1)^{j_y})\phi}, \\
t_{\bf j}^y & = & 2J e^{-i(-1)^{j_x+j_y}\phi}, \\
\Delta_x & = & 2J, \\
\Delta_y & = & r,
\end{array}
\ee
with $J=J_x=J_y$, $r = \sqrt{(2J)^2 +(4K)^2}$ and $\phi = \arctan (J/(2K))$. In terms of these variables the local phase $\theta_{\bf j}$  in \rf{Maj_pair}, that is chosen such that the pairing potentials $\Delta_x$ and $\Delta_y$ are real, is given by $\theta_{\bf j} = -(-1)^{j_x} {1-(-1)^{j_y} \over 2} \phi$. 

\begin{figure}[t]
\includegraphics[width=9cm]{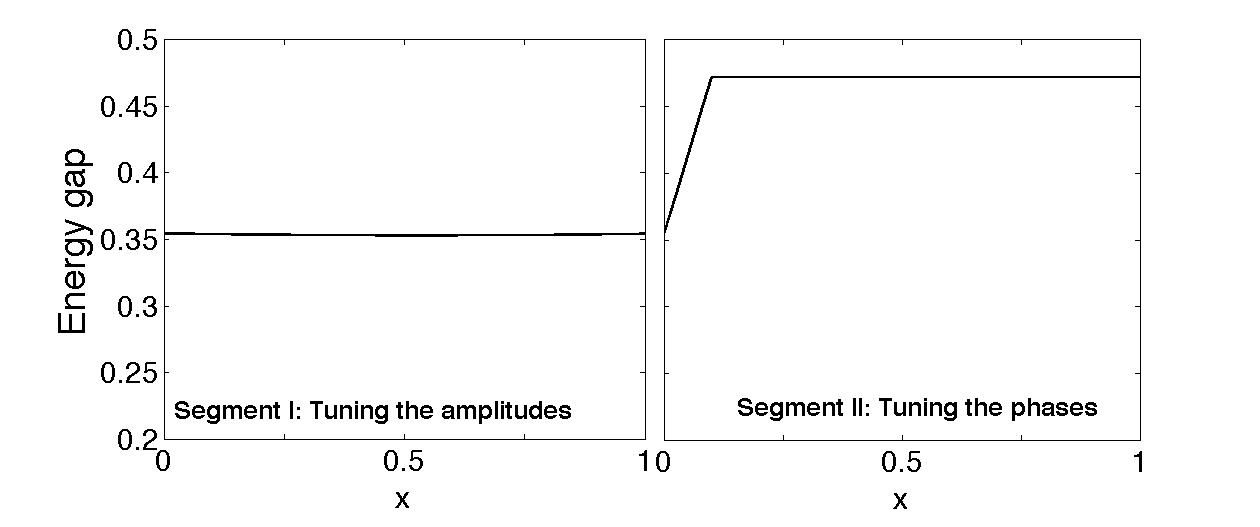}
\caption{\label{adcon} Adiabatic connection between the $\nu=2$ phases in the full-vortex sector of the honeycomb model and in \rf{H2}. We set $J_z=J=1$ and $K=-0.1$, which through the identifications \rf{honey_hopping} give the chemical potential $\delta=2$, $\mu=0$, while for the coupling amplitudes we get $\Delta_x=|t^y_j|=2$ and $\Delta_y=|t^x_j|=\sqrt{4.16} \approx 2.04$. {\it Left:} In the first segment of the adiabatic path we tune $t=\Delta_y=|t^x_j| \to \Delta_x=|t^y_j|$ to equalise all the amplitudes. The plot shows the energy gap of $H_\text{f.v.}[t(x)]$, where $t(x)=(1-x)\sqrt{4.16}+2x$, increasing monotonously during the process. {\it Right:} At the second step we tune the tunnelling phase $\phi=\textrm{arctan}(5) \to \pi/4$. The plot shows the energy gap of $H_\text{f.v.}[\phi(x)]$, where $\phi(x)=(1-x)\textrm{arctan}(5)+x\frac{\pi}{4}$, again first increasing and then settling to a constant value. Both transitions are performed with a linear ramp parameterised by $x\in[0,1]$.}
\end{figure}

The variables $t_{\bf j}^x$, $t_{\bf j}^y$, $r$ and $\phi$ are all functions of the honeycomb couplings $J$, $J_z$ and $K$. From now on we will treat them as independent variables and show that \rf{H_honey_BCS} can be adiabatically connected to \rf{H2}. We do this by explicitly constructing a path in the parameter space along which the energy gap remains finite. Due to the periodically alternating signs in the chemical potential $\mu_{\bf j}$, we begin with identifying the detuning $\delta$ with $2J_z$, where the overall chemical potential is   set to $\mu=0$. The first segment of the adiabatic path consists of tuning $r \to 2J \equiv t$, which makes both the tunnelling and pairing amplitudes equal ($|t_{\bf j}^x|=|t_{\bf j}^y|=\Delta_x=\Delta_y=t$). Fig. \ref{adcon} shows the gap essentially remaining constant during this process. 

At the second segment we tune the phases of $t_{\bf j}^x$ and $t_{\bf j}^y$ to match those of \rf{H2}. Writing out the tunnelling terms explicitly, we find the periodic pattern to consist of six independent ones, which are unitarily equivalent to
\be \label{coupling_deform}
\begin{array}{rclcl}
t^x_1 & = & t e^{i\pi/2} & \to & it, \\
t^x_2 & = & t e^{-i\pi/2} & \to & -it , \\
t^x_3 & = & t e^{4i\phi-i\pi/2} & \to & it , \\
t^x_4 & = & t e^{-4i\phi+i\pi/2} & \to & -it , \\
t^y_1 & = & t e^{-2i\phi+i\pi/2} & \to & t , \\
t^y_2 & = & t e^{2i\phi-i\pi/2} & \to & t,
\end{array}
\ee
as illustrated in Fig. \ref{mapping}. The arrow denotes the second segment of the adiabatic path where we tune $\phi \to \pi/4$ to make the tunnelling phases match those of \rf{H2}. Fig. \ref{adcon} shows the gap remaining again robust, which implies that our staggered model of spinless fermions is adiabatically connected to Kitaev's honeycomb model in the full-vortex sector. Indeed, for equal couplings $J=J_z=1$ and $K<0$ the honeycomb model is known to be in a $\nu=2$ phase \cite{Lahtinen10, Lahtinen12}. These honeycomb couplings correspond to $t=\Delta_x=\Delta_y=\delta=2$ and $\mu=0$ for which, as shown in Fig. \ref{PD}, we also find a $\nu=2$ phase. 

The phase diagram of the full-vortex sector of the honeycomb model has been studied in \cite{Lahtinen10}. The adiabatic connection between the models enables us to understand some of the features of the phase diagram of our model. First, we showed above that the full-vortex sector with equal couplings $|J_x|=|J_y|=|J_z|$ can be mapped onto the $\mu=0$, $\delta > 0$ line of Fig. \ref{PD}. Thus we can immediately understand the $\nu=2$ phase to correspond to the $\nu=2$ phase in the honeycomb model that is known to arise as the unique collective state of the Majorana modes bound to the vortex cores \cite{Lahtinen12}. When the staggering $\delta=2J_z$ of the hopping in \rf{Hfv} is gradually suppressed by introducing a finite $\mu$ by hand, it has been shown that for $\mu \gtrsim \delta/2$ the non-Abelian $\nu=1$ phase is recovered, even if some sign staggering remains. This is in agreement with Fig. \ref{PD}, which shows along the $\mu \approx \delta/2$ line a similar transition between the Abelian $\nu=2$ and the non-Abelian $\nu=-1$ phases (the change in the sign of the Chern number does not occur in the honeycomb model, but due to adiabatic deformation we expect only qualitatively similar behaviour in our model). The $\mu>\delta/2$ region of our model is thus adiabatically connected to the non-Abelian phase in the vortex-free sector (which in turn is adiabatically connected to the weak-pairing phase of a $p$-wave superconductor \cite{Yu08}). If isolated vortices were introduced there, they would bind localised Majorana modes with short range interactions \cite{Lahtinen11}. Finally, in the dimerised limits $|J_z| \gg |J|$ one should always find a $\nu=0$ phase, that corresponds to the strong pairing phase in $p$-wave superconductors. Indeed, Fig.~\ref{PD} shows a $\nu=0$ phase emerging in both $\delta \gg \mu$ and $\mu \gg \delta$ limits.

\section{Quantitative analysis of the optical lattice parameters}
\label{App_implementation}

We now provide a quantitative analysis to justify the feasibility of the model implementation in this work. Together with our own numerical simulations, we rely on the analysis provided in Refs.~\cite{Gerbier10,AllStars,Holland01}. We particularise our results to two interpenetrated square lattices, each of them with lattice constant $d \simeq 400$nm and hosting a hyperfine state of $^{40}$K.

The energy scale of the model parameters is constrained to an interval which depends on the lattice depth. This interval is bounded from below by the heating rates and the suppressed natural hopping within sublattices; it is bounded from above by the separation between lattice bands. We will show that all model parameters fit within this energy scale window, demonstrating the feasibility of the proposed implementation.

\begin{figure}[t]
\includegraphics[width=0.75\linewidth]{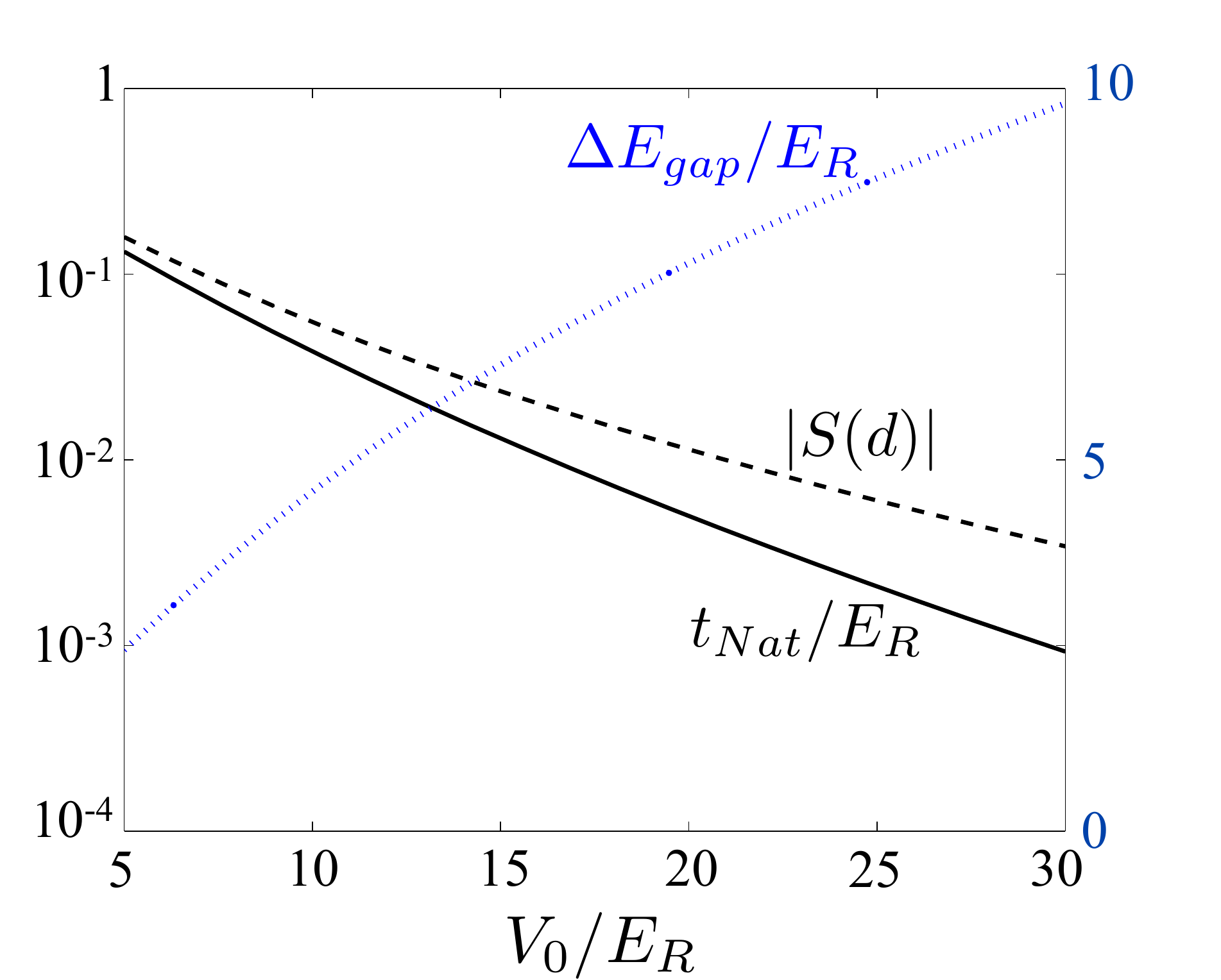}
\caption{Realistic band structure simulation of the intertwined lattice setup. \textit{Left axis:} Dependence of the natural hopping $t_{Nat}$ (black, solid) and the overlap between neighbouring wave functions $|S({d})|$ (black, dashed) on $V_0/E_R$. \textit{Right axis:} Dependence of the bandgap $\Delta E_{gap}$ (blue, dotted) on the lattice depth, $V_0/E_R$.} \label{SimQuant}
\end{figure}

We first focus on the lower end of this interval. The natural hopping parameter decreases roughly exponentially with the lattice depth. Our numerical simulations (Fig.~\ref{SimQuant}) show that the hopping reaches a value of $t_{Nat} \lesssim 10^{-3} E_R$ for a lattice with depth $V_0 \simeq 22E_R$,  where $V_0$ is the lattice depth and $E_R$ is the recoil energy of the lattice (around $E_R/h=8$kHz for the choice above). Therefore, we can expect a natural hopping of the order of $5$Hz. Our results are in agreement with previous analytical estimates~\cite{Zwerger}.

The second constraint lower bound of our interval of acceptable parameters is provided by the photon scattering rate, which increases with the depth the lattice. These heating rates are a significant problem for state-dependent setups, because in these designs the maximum detuning of light is limited by the energy splitting between hyperfine states. More precisely, the heating rate can be estimated as $\gamma_h\simeq (\Gamma/\delta_{Deph})V_0$, where $\Gamma$ is the spontaneous emission rate of the atom, and $\delta_{Deph}$ the detuning. The ratio $\Gamma/\delta_{Deph}$ critically depends on the atomic species, ranging from about $0.1/h$ for $^6Li$ to about $10^{-5}/h$ for $^{40}K$. We focus on this last atomic element, obtaining a heating rate of about $1-2$ Hz for the above mentioned $V_0 \sim 22 E_R$, but we remark the possibility of using alkaline-earth atoms to bring this value down to about $0.01$ Hz~\cite{Gerbier10}.

Finally, all energy scales must be significantly smaller than the bandgap, $\Delta E_{gap}$. Our simulations evaluate this bandgap to be over $60$kHz for our $V_0 \simeq 22E_R$ lattice (Fig.~\ref{SimQuant}). Again this result agrees with comparable calculations in similar setups~\cite{Gerbier10}. In summary, our parameters ($\mu_b-\mu_w,t,\Delta$) should all move in the $0.1-1$kHz range in order to successfully implement our proposed model.

The chemical potential difference $\delta$ can independently tuned by the auxiliary offset lattice intensity $V_{OFF}$. This offset can be easily set to the desired energy range, since it just requires a superlattice modulation which is much smaller than the intensity of the main lattice ($V_{OFF}<E_R$).

The Raman tunnelling $t$ is proportional to the Raman beam intensity, $|t|=\hbar \Omega S(d)$, and the overlap between Wannier wave functions, $w(x,y)$, in neighboring wells of the superlattice, $S({d})=\iint w^{\star}(x,y)  w(x-d/2,y-d/2) \mathrm{d}x\mathrm{d}y$. We estimate numerically this overlap to be $S \simeq 10^{-2}$ for $V_0=22 E_R$ (Fig.~\ref{SimQuant}). Therefore, a feasible value $\Omega\sim E_R/\hbar$ would keep $|t|$ in the desired 1 kHz order of magnitude.

Finally, the pairing $\Delta$ depends on the strength of the coupling to the molecular reservoir and the bosonic bath density~\cite{Holland01} as $|\Delta| =g \sqrt{\rho} S({d})$. The overlap of the fermionic wave functions again plays an important role and we assume the density profile of the bosonic bath to be uniform. Estimates from previous proposals~\cite{AllStars} based on condensed fermionic pair experiments~\cite{Zwierlein04} show that $|\Delta|\simeq 0.5$kHz is challenging but possible between nearest neighbours of the superlattice.

\end{document}